\documentclass[%
preprint,
 amsmath,
 amssymb,
 prfluids,
]{revtex4-2}

\usepackage{graphicx}% Include figure files
\usepackage{dcolumn}% Align table columns on decimal point
\usepackage{bm}% bold math
\usepackage{scalerel}
\usepackage{stackengine,wasysym}

\newcommand\reallywidetilde[1]{\ThisStyle{%
  \setbox0=\hbox{$\SavedStyle#1$}%
  \stackengine{-.1\LMpt}{$\SavedStyle#1$}{%
    \stretchto{\scaleto{\SavedStyle\mkern.2mu\AC}{.5150\wd0}}{.5\ht0}%
  }{O}{c}{F}{T}{S}%
}}

\usepackage{xcolor}

\begin{document}

%\preprint{APS/123-QED} 

\title{A local dynamic gradient Smagorinsky model for large-eddy simulation}% Force line breaks with \\
\author{Wybe Rozema}
\email{w.rozema@protonmail.com}
\affiliation{Independent researcher, Diemen, The Netherlands}

\author{H. Jane Bae} 
\email{jbae@caltech.edu}
\affiliation{Graduate Aerospace Laboratories, California Institute of Technology, Pasadena, CA 91125}

\author{Roel W. C. P. Verstappen} 
\email{r.w.c.p.verstappen@rug.nl} 
\affiliation{Bernoulli Institute for Mathematics, Computer Science and Artificial Intelligence, University of Groningen, Nijenborgh 9, 9747 AG Groningen, The Netherlands}

%\affiliation{
%\affiliation{
%\affiliation{%


\date{\today}% It is always \today, today,

\begin{abstract}
This paper introduces a local dynamic model for large-eddy simulation (LES) without averaging in the homogeneous directions. It is demonstrated that the widely-used dynamic Smagorinsky model (DSM) has a singular dynamic model constant if it is used without averaging. The singularity can cause exceedingly large local values of the dynamic model constant. If these large values are not mitigated by the application of averaging, they can amplify discretization errors and impair the stability of simulations. To improve the local applicability of the DSM, the singularity is removed by replacing the resolved rate-of-strain tensors in the Smagorinsky model with the resolved velocity gradient tensor. These replacements result in the new dynamic gradient Smagorinsky model (DGSM). Results of simulations of three canonical turbulent flows (decaying homogeneous isotropic turbulence, a temporal mixing layer, and turbulent channel flow) are presented to demonstrate the potential of this model. The DGSM provides improved stability compared to the local DSM and does not require averaging for stability at time step sizes that are typically used for a locally consistent static LES model. Results obtained with the DGSM are generally as accurate as results obtained with the DSM, while the DGSM has lower computational complexity. Moreover, the DGSM is easy to implement and does not require any homogeneous direction in space or time. It is therefore concluded that the DGSM is a promising local dynamic model for LES.
%\begin{description}
%\end{description}
\end{abstract}

\maketitle

%\tableofcontents

\section{\label{sec:introduction}Introduction}
Understanding turbulent flow is important for many engineering and environmental challenges such as increasing the power output of wind farms \cite{sorensen-2011,stevens-2017}, reducing the fuel consumption and emissions of aircraft \cite{slotnick-2014}, and predicting atmospheric flows \cite{stoll-2020}. However, turbulent flow often cannot be simulated by direct numerical simulation (DNS) because the computational complexity of resolving the small turbulent flow structures is excessive \cite{pope-2000,choi-2012,yang-2021}. Therefore, simulations of turbulent flow are typically based on approximate models which can be solved at lower computational costs. An example of an approximate model for turbulent flow is LES. In LES, the large energy-carrying eddies in a flow are simulated directly, whereas the effect of small unresolved sub-filter scales on the large eddies is modeled. 

Many LES models are based on an eddy viscosity. Eddy-viscosity models assume that the character of the sub-filter scales is dissipative and that the dissipation rate of resolved kinetic energy is determined by an eddy viscosity which is modeled based on the resolved flow. A traditional eddy-viscosity model is the Smagorinsky model \cite{smagorinsky-1963}. Although adequate results can be obtained with the Smagorinsky model in simulations of decaying homogeneous isotropic turbulence, the model provides excessive sub-filter dissipation for other flows. For simulations of turbulent channel flow, the model constant of the Smagorinsky model should be decreased and a wall-damping function should be applied to obtain adequate results \cite{deardorff-1971,moin-1982}, and for simulations of a temporal mixing layer the Smagorinsky model incorrectly suppresses transition to turbulence \cite{vreman-1997}. 

The excessive sub-filter dissipation of the Smagorinsky model can be corrected by the derivation of more appropriate models for the eddy viscosity. Such LES models are often derived by imposing properties of the eddy viscosity which are expected to be desirable \cite{vreman-2004,kobayashi-2005,nicoud-2011,rozema-2015}. The advantages of these models are that they can be straightforwardly implemented in a simulation method and that their computational complexity is relatively low. A perceived disadvantage is that the models include at least one model constant which requires calibration. The appropriate value of the model constant can depend on both the used simulation method and the simulated flow \cite{verstappen-2014,park-2006}. The requirement to calibrate the model constant limits the ease of use and range of applicability of LES models with a more appropriate model for the eddy viscosity.

An alternative correction of the excessive sub-filter dissipation of the Smagorinsky model is to maintain the unsatisfactory model for the eddy viscosity but to correct for the shortcomings of this eddy viscosity by adjusting the model constant using the dynamic procedure. The dynamic procedure is based on a mathematical identity of sub-filter scale terms for different filter widths which is known as the Germano identity \cite{germano-1991}. By applying a test filter to the LES solution and solving the Germano identity by averaging in the directions of statistical homogeneity, the model constant for the Smagorinsky model can be determined dynamically based on the simulated flow. Using the obtained averaged dynamic model constant in the Smagorinsky model gives the DSM \cite{germano-1991,lilly-1992}. Satisfactory results can be obtained with the DSM for simulations of flows with directions of statistical homogeneity such as turbulent channel flow and a temporal mixing layer \cite{germano-1991,piomelli-1993,vreman-1997}. If the dynamic model constant of the DSM is determined locally, it can attain excessive values which can cause instability of simulations \cite{germano-1991,kobayashi-2005}. Therefore, although the Germano identity holds locally, the DSM is typically not applied locally. Instead, it is common practice to apply averaging of the dynamic model constant of the DSM in directions of statistical homogeneity \cite{germano-1991,pope-2000}.

Despite the satisfactory results obtained for flows with directions of statistical homogeneity, the application of averaging is typically perceived to be a disadvantage of the DSM. The application of averaging increases the computational complexity of simulations. Also, averaging can make the DSM less dynamic, because the obtained model constant does not only depend on the local flow. Finally, evident directions of statistical homogeneity do not always exist for complex geometries, and therefore the application of averaging is not always straightforward. These disadvantages reduce the range of applicability of the DSM.

Commonly used solutions to overcome the latter disadvantage are to solve the dynamic model constant using a constrained variational formulation \cite{ghosal-1995} or to apply weighted averaging over flow-path lines of fluid particles \cite{meneveau-1996}. These solutions extend the range of applicability of the DSM to complex geometries, but they do not fully address all the disadvantages related to averaging. An alternative practical solution is to combine local averaging of the dynamic model constant and clipping of negative and large values \cite{noma-2007,chang-2019}. In some implementations of the DSM the dynamic constant is locally averaged by application of the test filter and negative values and values larger than a threshold are clipped. A disadvantage of combining local averaging and clipping is that the domain used for the local averaging and the clipping thresholds are not supported theoretically and that it is not evident that the applied treatments improve the accuracy of LES results.

This study investigates the instability of the DSM and proposes an alternative solution to overcome the instability without averaging. This results in a local dynamic LES model with improved stability. The governing equations for LES are introduced in Sec.~\ref{sec:LES}. The instability of the DSM is investigated in Sec.~\ref{sec:DSM}. Based on the identified source of the instability, a new dynamic model based on a modified Smagorinsky model is proposed in Sec.~\ref{sec:LDMs}. The stability and accuracy of the proposed model are assessed for simulations of decaying homogeneous isotropic turbulence, a temporal mixing layer, and turbulent channel flow in Sec.~\ref{sec:results}.

\section{\label{sec:LES}Large-eddy simulation}
The governing equations for LES can be obtained by the application of a spatial LES filter to the incompressible Navier-Stokes equations \cite{leonard-1975,pope-2000}. The spatial LES filter is typically related to the computational grid used for simulations. If the LES filter is assumed to commute with spatial derivation, this gives the LES equations
\begin{equation}
\label{eq:leseqs}
\partial_t \overline{u}_i + \partial_j \left(\overline{u}_i \overline{u}_j \right)  + \partial_i \overline{p}  - \partial_j \left( \nu  \partial_j \overline{u}_i \right) = - \partial_j \tau_{i j} (\bm{u}) \; , \quad \partial_i \overline{u}_i = 0 \; ,
\end{equation}
where $\overline{u}_i$ is the LES-filtered flow velocity in the direction $x_{i}$, $\overline{p}$ is the LES-filtered pressure, $\nu$ is the kinematic viscosity, $\tau_{i j}(\bm{u}) = \overline{u_i u_j} - \overline{u}_i \overline{u}_j$ is the sub-filter tensor, and the Einstein summation convention is used. The sub-filter tensor represents the effect of the sub-filter scales on the resolved flow. By multiplication of the above equation by $\overline{u}_i$ and application of partial integration, it can be shown that the local sub-filter dissipation of the exact sub-filter tensor is equal to
\begin{equation}
\label{eq:exactsgsdiss}
\varepsilon_{\tau} = - \tau_{i j} (\bm{u}) \left( \partial_j \overline{u}_i \right) \; .
\end{equation}
LES requires approximation of the effect of sub-filter scales by closure the filtered Navier-Stokes equations with a model for the tensor $\tau_{i j} (\bm{u})$ based on the LES-filtered flow velocity $\overline{\bm{u}}$. The objective of LES modeling is to derive a sub-filter model $\tau_{i j} (\overline{\bm{u}})$ for which solutions of the LES equations in Eq.~(\ref{eq:leseqs}) accurately approximate filtered solutions of the Navier-Stokes equations. The local sub-filter dissipation provided by an LES model is equal to
\begin{equation}
\label{eq:modelsgsdiss}
\varepsilon_{m} = - \tau_{i j} (\overline{\bm{u}}) \left( \partial_j \overline{u}_i \right) \; .
\end{equation}

\section{\label{sec:DSM}The dynamic Smagorinsky model}
A traditional LES model is the Smagorinsky model \cite{smagorinsky-1963}. The Smagorinsky model assumes that the anisotropic part of the sub-filter model tensor is equal to
\begin{equation}
\label{eq:smagorinsky}
\tau_{i j} (\overline{\bm{u}}) - \frac{1}{3} \tau_{k k} (\overline{\bm{u}}) {\delta}_{i j} = - 2 C_{S} \overline{\Delta}^2  \left\lvert \overline{S} \right\rvert \overline{S}_{i j} \; ,
\end{equation}
where $\delta$ is the identity tensor, $C_{S}$ is the model constant, $\overline{\Delta}$ is the width of the LES filter, $\overline{S}_{i j} = ( \partial_i \overline{u}_j + \partial_j \overline{u}_i )/2$ is the resolved rate-of-strain tensor, and $\left\lvert \overline{S} \right\rvert = \sqrt{2 \overline{S}_{i j} \overline{S}_{i j} }$. In this paper it is assumed that the width of the LES filter is approximated by $\overline{\Delta}^3 = \overline{\Delta}_{1} \overline{\Delta}_{2} \overline{\Delta}_{3} $, where $\overline{\Delta}_i$ is the width of the LES filter in the direction $x_i$ \cite{deardorff-1970}. The sub-filter dissipation provided by the Smagorinsky model is
\begin{equation}
\label{eq:smagorinsky_diss}
\varepsilon_{m} = - \tau_{i j} (\overline{\bm{u}}) \left( \partial_j \overline{u}_i \right)  = C_{S} \overline{\Delta}^2  \left\lvert \overline{S} \right\rvert^3 \; .
\end{equation}
Accurate results in simulations of decaying homogeneous isotropic turbulence are obtained with the Smagorinsky model if the square root of the model constant $C_{S}$ is set to approximately $0.17$ \cite{pope-2000}. 
However, this value of the model constant is too large to obtain adequate results for general flows. A smaller model constant is required to obtain adequate results with the Smagorinsky model in simulations of turbulent channel flow and a temporal mixing layer \cite{deardorff-1971,moin-1982,vreman-1995}.  

The applicability of the Smagorinsky model for general flows can be improved by the application of the dynamic procedure \cite{germano-1991}. The dynamic procedure calculates the value of the model constant dynamically by comparing the filtered flow velocity and the LES model for two different filter widths. The dynamic procedure applies a test filter with filter width $\widetilde{\Delta}_i$ to the filtered flow velocity $\overline{\bm{u}}$. This results in the test-filtered resolved velocity $\widetilde{\overline{\bm{u}}}$, which satisfies the LES equations for subsequent application of the LES and test filters. The dynamic procedure is based on the Germano identity
\begin{equation}
\label{eq:germano-identity}
L_{i j} (\overline{\bm{u}}) = T_{i j}(\bm{u}) - \widetilde{\tau_{i j}(\bm{u}) } \; ,
\end{equation}
where 
\begin{equation}
\label{eq:leonardtens}
L_{i j} (\overline{\bm{u}}) = \widetilde{\overline{u}_i \overline{u}_j} - \widetilde{\overline{u}}_i \widetilde{\overline{u}}_j
\end{equation}
is the Leonard tensor which can be calculated in an LES by approximation of the test filter, and $T_{i j}(\bm{u})= \widetilde{\overline{u_i u_j}} - \widetilde{\overline{u}}_i \widetilde{\overline{u}}_j$ is the sub-filter tensor for subsequent application of the LES and test filters. Application of the Smagorinsky model at the aggregated filter level gives
\begin{equation}
\label{eq:smagorinsky_test}
T_{i j} (\overline{\bm{u}}) - \frac{1}{3} T_{k k} (\overline{\bm{u}}) {\delta}_{i j} = - 2 C_{S} \widetilde{\overline{\Delta}}^2  \left\lvert \widetilde{\overline{S}} \right\rvert \widetilde{\overline{S}}_{i j} \; .
\end{equation}
Substitution of the Smagorinsky model in the Germano identity and assuming that the model constant and LES filter width do not change over the width of the test filter gives
\begin{equation}
\label{eq:germano-identity_dsm}
L_{i j} (\overline{\bm{u}}) - \frac{1}{3} L_{k k} (\overline{\bm{u}}) {\delta}_{i j} = T_{i j}(\overline{\bm{u}}) - \widetilde{\tau_{i j}(\overline{\bm{u}})} = C_{S} M_{i j} (\overline{\bm{u}}) \; ,
\end{equation}
where
\begin{equation}
\label{eq:germano-identity_dsmm}
M_{i j} (\overline{\bm{u}}) = - 2 \widetilde{\overline{\Delta}}^2  \left\lvert \widetilde{\overline{S}} \right\rvert \widetilde{\overline{S}}_{i j} + 2   \overline{\Delta}^2 \widetilde{ \left\lvert \overline{S} \right\rvert \overline{S}_{i j} } \; ,
\end{equation}
which can be calculated in an LES by approximation of the test filter. %Although the assumption that the model constant and filter width are constant made to obtain the above equation is not expected to be valid in general, making these assumptions for the purpose of deriving the dynamic model constant is common practice in literature \cite{germano-1991,lilly-1992}. 

The dynamic model constant can be obtained by minimizing the error of the Germano identity in Eq.~(\ref{eq:germano-identity_dsm}) using a least-squares approach \cite{lilly-1992}. This gives the model constant
\begin{equation}
\label{eq:model_constant_dsm}
C_{S} = \frac{L_{i j} M_{i j}}{M_{k l} M_{k l}} \; .
\end{equation}
If the tensor $M_{i j}$ in Eq.~(\ref{eq:germano-identity_dsmm}) vanishes, then any value of $C_{S}$ corresponds to a least-squares solution and therefore the solution $C_{S} = 0$ can be used. Substitution of this dynamically determined model constant in the Smagorinsky model in Eq.~(\ref{eq:smagorinsky}) gives the DSM before any application of averaging or clipping. In simulations this locally determined model constant can attain very large values and negative values, which can cause numerical instability of simulations \cite{pope-2000}. This instability is typically treated by averaging the numerator and denominator of the least-squares solution and clipping negative values of the obtained model constant to zero
\begin{equation}
\label{eq:model_constant_dsm_av}
C_{S} = \max \left\{ \frac{ \langle L_{i j} M_{i j} \rangle }{\langle M_{k l} M_{k l} \rangle } , 0 \right\} \; ,
\end{equation}
where the brackets denote averaging. The averaging is typically applied in directions of statistical homogeneity \cite{germano-1991}. In this paper, the DSM with the above dynamic model constant is called the averaged DSM.

To demonstrate the instability of simulations with the DSM if averaging is not applied, this paper considers also the dynamic model constant without averaging, but with clipping negative values of the model constant to zero
\begin{equation}
\label{eq:model_constant_dsm_local}
C_{S} = \max \left\{ \frac{ L_{i j} M_{i j} }{M_{k l} M_{k l}} , 0 \right\} \; .
\end{equation}
The DSM with this dynamic constant without averaging is called the local DSM.

Adequate LES results have been obtained with the averaged DSM for turbulent channel flow, the temporal mixing layer, and other flows \cite{germano-1991,piomelli-1993,vreman-1997,meneveau-2000}. However, the need to apply averaging to prevent instability of simulations can be perceived to be a shortcoming of the DSM, because it increases the computational complexity of simulations, it is not straightforward to apply appropriate averaging for simulations of the flow around complex geometries, and because the averaging can result in the application of sub-filter dissipation in regions of laminar flow. The need for averaging also does not seem to follow from the first principles of the dynamic procedure because the Germano identity in Eq.~(\ref{eq:germano-identity}) holds locally. This motivates investigation of the source of the instability and alternative approaches to overcome or mitigate the instability.

\subsection{\label{sec:singularityDSM}Singularity of the model constant of the dynamic Smagorinsky model}
In this section, it is demonstrated that the model constant of the DSM has a singularity, which can cause exceedingly large values of the model constant locally. The singularity is derived for the model constant of the DSM in Eq.~(\ref{eq:model_constant_dsm}) without application of averaging or clipping negative values of the model constant to zero. For the purpose of the derivation of the singularity, the LES filter and test filter are assumed to be a box filter. The filter width of the test filter is commonly set equal to a multiple of the filter width of the LES filter, such that $\widetilde{\Delta}_i = \beta \overline{\Delta}_i$ for some $\beta$. Subsequent application of box filters with filter widths $\overline{\Delta}_{i}$ and $\widetilde{\Delta}_{i}$ is not equivalent to application of a box filter with an effective filter width. However, the resulting aggregated filter is most accurately approximated by a box filter with filter width $\widetilde{\overline{\Delta}}_{i}^2 = \widetilde{\Delta}_{i}^2 + \overline{\Delta}_{i}^2$ \cite{vreman-1994}. For the purpose of the derivation of the singularity, the effective filter width of the aggregated filter in each direction is therefore approximated by $\widetilde{\overline{\Delta}}_{i}^2 = \left(1+ {\beta}^2 \right) \overline{\Delta}_{i}^2$. Thus, the filter width used in the Smagorinsky model at the aggregated filter level satisfies $\widetilde{\overline{\Delta}}^2 = \left( 1+{\beta}^2 \right) \overline{\Delta}^2$.

The leading-order behavior of the singularity can be analyzed using series expansions of test-filtered quantities. A series expansion of a test-filtered quantity $f$ can be obtained by applying the test filter to a Taylor expansion of $f$, which gives
\begin{equation}
\label{eq:taylortestfilter}
\widetilde{f} =  f + \frac{1}{24} \widetilde{\Delta}_i^2 \partial^{2}_{i} f +{O} (\widetilde{\Delta}^4) \; .
\end{equation}
This series expansion exists if the test-filtered quantity is infinitely differentiable \cite{vreman-1996}. Application of this series expansion to the Leonard tensor gives
\begin{equation}
\label{eq:leonard_taylor}
L_{i j} (\overline{\bm{u}}) = \frac{1}{12}  \widetilde{\Delta}_k^2 \left( \partial_k \overline{u}_i \right) \left(\partial_k \overline{u}_j \right) + {O} (\widetilde{\Delta}^4) = \beta^2 \frac{1}{12}  \overline{\Delta}_k^2\left( \partial_k \overline{u}_i \right) \left(\partial_k \overline{u}_j \right)  + {O} (\overline{\Delta}^4) \; .
\end{equation}
The leading-order term of this series expansion is proportional to the leading-order term of the series expansion of the exact sub-filter tensor \cite{clark-1979,vreman-1996}
\begin{equation}
\label{eq:clark}
\tau_{i j} (\bm{u}) = \frac{1}{12}  \overline{\Delta}_k^2 \left( \partial_k \overline{u}_i \right) \left(\partial_k \overline{u}_j \right) + {O} (\overline{\Delta}^4) \; .
\end{equation}
Derivation of the series expansion of the tensor in Eq.~(\ref{eq:germano-identity_dsmm}), and substitution of the relation of the filter widths of the LES and test filters gives
\begin{equation}
\label{eq:germano-identity_dsmm_taylor}
M_{i j} (\overline{\bm{u}}) = - 2 \beta^2 \overline{\Delta}^2 \left\lvert \overline{S} \right\rvert \overline{S}_{i j}  + {O} (\overline{\Delta}^4) \; .
\end{equation}
Substitution of the series expansions in Eqs.~(\ref{eq:leonard_taylor}) and (\ref{eq:germano-identity_dsmm_taylor}) in Eq.~(\ref{eq:model_constant_dsm}) gives a series expansion of the dynamic model constant \cite{pope-2000}
\begin{equation}
\label{eq:cs_taylor}
C_{S} = - \frac{1}{12} \frac{ \overline{\Delta}_k^2 \left( \partial_k \overline{u}_i \right) \left(\partial_k \overline{u}_j \right) \overline{S}_{i j}}{\overline{\Delta}^2 \left\lvert \overline{S} \right\rvert^3} + {O} (\overline{\Delta}^2) \; .
\end{equation}
An alternative series expansion of the dynamic model constant based on test-filtered quantities can be obtained by substitution of Eq.~(\ref{eq:taylortestfilter}) in the above series expansion
\begin{equation}
\label{eq:cs_taylor_tf}
C_{S} = - \frac{1}{12} \frac{ \overline{\Delta}_k^2 \left( \partial_k \widetilde{\overline{u}}_i \right) \left(\partial_k \widetilde{\overline{u}}_j \right) \widetilde{\overline{S}}_{i j}}{\overline{\Delta}^2 \left\lvert \widetilde{\overline{S}} \right\rvert^3} + {O} (\overline{\Delta}^2) \; .
\end{equation}
The above series expansions will be used to analyze the singularity of the model constant of the DSM.

The series expansions can also be used to further understand the DSM and its relation to other LES models. Substitution of the series expansion in Eq.~(\ref{eq:cs_taylor}) in the Smagorinsky model in Eq.~(\ref{eq:smagorinsky_diss}) and assuming that the resolved rate-of-strain tensor does not vanish, gives a series expansion of the dissipation provided by the model constant of the local DSM in Eq.~(\ref{eq:model_constant_dsm})
\begin{equation}
\label{eq:dsmdiss}
\varepsilon_{m} = - \frac{1}{12}  \overline{\Delta}_k^2 \left( \partial_k \overline{u}_i \right) \left(\partial_k \overline{u}_j \right) \overline{S}_{i j}+ {O} (\overline{\Delta}^4) \; .
\end{equation}
The leading-order term of this series expansion is equal to the leading-order term of the series expansion of the exact sub-filter dissipation based on Eqs.~(\ref{eq:exactsgsdiss}) and (\ref{eq:clark}). Therefore, the model constant of the DSM without averaging and clipping approximates the exact sub-filter dissipation. The leading-order consistency of the sub-filter dissipation is a strength of the DSM that has been identified previously, although the provided arguments are not based on a comparison of leading-order terms \cite{jimenez-1995}. Because the model constant of the DSM is locally consistent with the exact sub-filter dissipation, the application of spatial averaging can weaken this local consistency. The sub-filter dissipation of the DSM without averaging in Eq.~(\ref{eq:dsmdiss}) is included in the numerator of the anisotropic minimum-dissipation (AMD) model for LES \cite{rozema-2015}. This could explain why simulations with the AMD model and the averaged DSM give similar results for many flows \cite{rozema-2015,abkar-2018,gadde-2021}.

The singularity of the model constant of the DSM can be understood conceptually by comparing the numerator and denominator of the leading-order term of its series expansions in Eqs.~(\ref{eq:cs_taylor}) and (\ref{eq:cs_taylor_tf}). For both the series expansions, the numerator is proportional to a contraction of the (test-filtered) resolved velocity gradient and rate-of-strain tensors, whereas the denominator is proportional to a power of a contraction of only the (test-filtered) resolved rate-of-strain tensor. If the resolved rate-of-strain tensor vanishes, but the resolved velocity gradient tensor is non-zero because it is primarily determined by a non-zero resolved rate-of-rotation tensor $\overline{\Omega}_{i j} = \partial_{i} \overline{u}_{j} - \overline{S}_{i j}$, then the denominator vanishes at a higher rate than the numerator. Therefore the leading-order term of the series expansion of the model constant of the DSM has a singularity and can attain excessive values.

To mathematically derive the singularity, assume that at a location $\bm{x}_0$ the filtered velocity behaves linearly for a domain larger than the filter width of the test filter
\begin{equation}
\overline{u}_i = \overline{u}_i (\bm{x}_0) + A_{j i} \left(x_j - \left(x_0 \right)_j \right) =  u_i (\bm{x}_0) + A_{j i} \left(x_j - \left(x_0 \right)_j \right)  \; .
\end{equation}
Then the resolved velocity gradient tensor and the test-filtered resolved velocity gradient tensor are equal to the tensor $\partial_i \overline{u}_j = \partial_i \widetilde{\overline{u}}_j = A_{i j}$ at the location $\bm{x}_0$, and the dynamic model constant of the DSM is equal to the leading-order term of the series expansions in Eqs.~(\ref{eq:cs_taylor}) and (\ref{eq:cs_taylor_tf}). For the derivation of the singularity, the LES filter and test filter are assumed to be isotropic, which gives the dynamic model constant
\begin{equation}
\label{eq:cs_taylor_iso}
C_{S} = - \frac{1}{12} \frac{ \left( \partial_k \overline{u}_i \right) \left(\partial_k \overline{u}_j \right) \overline{S}_{i j}}{\left\lvert \overline{S} \right\rvert^3} = - \frac{1}{12} \frac{ \left( \partial_k \widetilde{\overline{u}}_i \right) \left(\partial_k \widetilde{\overline{u}}_j \right) \widetilde{\overline{S}}_{i j}}{\left\lvert \widetilde{\overline{S}} \right\rvert^3} \; .
\end{equation}
It is assumed that the locally linear flow is of the form
\begin{equation}
\label{eq:locallinearflow}
A_{i j} = \gamma \left\lvert A \right\rvert A^{S}_{i j} +  \sqrt{1 - \gamma^2} \left\lvert A \right\rvert A^{\Omega}_{i j} \; ,
\end{equation}
where $\gamma$ is a scalar between 0 and 1, $ \left\lvert A \right\rvert = \sqrt{2 A_{i j} A_{i j}}$ is assumed to be fixed, $A^{S}$ is a symmetric tensor with $\left\lvert A^{S} \right\rvert = 1$ and zero trace, and  $A^{\Omega}$ is an anti-symmetric tensor with $\left\lvert A^{\Omega} \right\rvert = 1$. The tensor $A_{i j}$ is symmetric if $\gamma$ is equal to 1, and anti-symmetric if $\gamma$ is equal to 0. The (test-filtered) resolved velocity gradient is bounded and $\left\lvert \nabla \overline{\bm{u}} \right\rvert = \sqrt{2 \partial_i \overline{u}_j  \partial_i \overline{u}_j }$ and $\left\lvert \nabla \widetilde{\overline{\bm{u}}} \right\rvert = \sqrt{2 \partial_i \widetilde{\overline{u}}_j  \partial_i \widetilde{\overline{u}}_j }$ are equal to the fixed value $ \left\lvert A \right\rvert$ for all values of $\gamma$. For the flow defined above, the exact (test-filtered) resolved rate-of-strain tensor is $\overline{S} = \widetilde{\overline{S}} = \gamma \left\lvert A \right\rvert A^{S} $ and therefore $\left\lvert \overline{S} \right\rvert = \left\lvert \widetilde{\overline{S}} \right\rvert = \gamma \left\lvert A \right\rvert = \gamma \left\lvert \nabla \overline{\bm{u}} \right\rvert = \gamma \left\lvert \nabla \widetilde{\overline{\bm{u}}} \right\rvert$.

The singularity can be demonstrated by setting the tensors $A^{S}$ and $A^{\Omega}$ equal to, for example
\begin{equation}
A^{S} = \frac{1}{2 \sqrt{2}} \begin{pmatrix}
0 & 1 & 0 \\
1 & 1 & 0 \\
0 & 0 & -1 
\end{pmatrix}  \; , \quad A^{\Omega}=  \frac{1}{2  \sqrt{2}}  \begin{pmatrix}
0 & 0 & 1 \\
0 & 0 & -1 \\
-1 & 1 & 0 
\end{pmatrix}  \; .
\end{equation}
Substitution of these tensors in Eq.~(\ref{eq:cs_taylor_iso}) gives the dynamic model constant
\begin{equation}
\label{eq:exactmodelconstant}
C_{S}  = \frac{1}{192 \sqrt{2}} \left( \frac{3}{\gamma^2} - 6 \right) \; ,
\end{equation}
which approaches infinity as $\gamma$ vanishes. Thus, for the considered locally linear flow in Eq.~(\ref{eq:locallinearflow}), the model constant of the DSM at $\bm{x}_0$ has a singularity at $\gamma = 0$, for which the (test-filtered) resolved rate-of-strain tensor vanishes while the (test-filtered) velocity gradient tensor is bounded and has a fixed norm equal to $\left\lvert A \right\rvert$. For small values of $\gamma$ the model constant is positive, and therefore clipping negative values to zero does not remove the identified singularity. The parameter $\gamma$ satisfies the equation $\left\lvert \overline{S} \right\rvert = \left\lvert \widetilde{\overline{S}} \right\rvert = \gamma \left\lvert \nabla \overline{\bm{u}} \right\rvert = \gamma \left\lvert \nabla \widetilde{\overline{\bm{u}}} \right\rvert$. This suggests that the leading-order behavior of the model constant of the DSM is $\left\lvert C_{S} \right\rvert \propto 1/ \left( \left\lvert \overline{S} \right\rvert^2 / \left\lvert \nabla \overline{\bm{u}} \right\rvert^2 \right)$ or $\left\lvert C_{S} \right\rvert \propto 1/ \left( \left\lvert \widetilde{\overline{S}} \right\rvert^2 / \left\lvert \nabla \widetilde{\overline{\bm{u}}} \right\rvert^2 \right)$ close to the singularity. 

% ================

To assess the relevance of the derived singular leading-order behavior for LES of turbulent flows, the model constant of the DSM is calculated based on results of a DNS of forced isotropic turbulence at a Taylor Reynolds number of $\textnormal{Re}_{\lambda} \sim 433$ in a cube of length $2 \pi$ with periodic boundary conditions \cite{perlman2007}. Results of the DNS are available for points on a grid with 1024 cells in each direction, and for different times after the simulation attains a statistically stationary state at $t=0$ \cite{Li2008}. The LES filter is set to a box filter with a filter width equal to $\overline{\Delta}_i = \pi / 16$ and the test filter is a box filter with a filter width equal to $\widetilde{\Delta}_i = 2 \overline{\Delta}_i$. The local model constant of the DSM in Eq.~(\ref{eq:model_constant_dsm}) is calculated accurately for points at cell centers of a grid with 64 cells in each direction at $t = 0$. The tensors in Eqs.~(\ref{eq:leonardtens}) and (\ref{eq:germano-identity_dsmm}) are calculated by discretization of the LES and test filters using midpoint integration, and by discretization of gradients using a central second-order finite difference method based on the cells of the computational grid of the DNS. Because the discretization of the gradients is applied to quantities that are filtered over at least 32 computational grid cells, the numerical discretization errors are expected to have a negligible impact on the calculation of the dynamic model constant. 

\begin{figure*}
\includegraphics{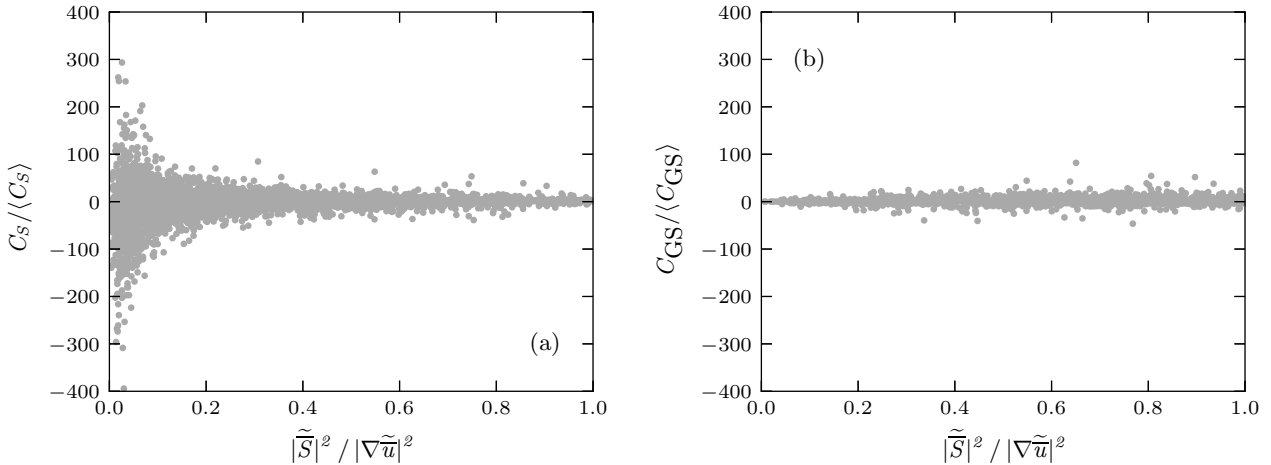}
\caption{\label{fig:fig_dsm_dgsm_singularity} Scatter plot of the local dynamic constant normalized by the sample mean calculated based on DNS results for the DSM (a) and the DGSM (b).}
\end{figure*}

The sample mean $\langle C_{S} \rangle$ of the calculated local values of the model constant of the DSM is equal to $1.38 \times 10^{-2}$. This value for a box filter is somewhat smaller than the theoretical average value for a spectral cutoff filter \cite{pope-2000}. % obtained average value is 0.013754094588483246 
A scatter plot of the normalized model constant of the DSM against the ratio $\left\lvert \widetilde{\overline{S}} \right\rvert^2 / \left\lvert \nabla \widetilde{\overline{\bm{u}}} \right\rvert^2$ is shown in Fig.~\ref{fig:fig_dsm_dgsm_singularity}(a). 
The occurrence of relatively large values of the model constant for vanishing $\left\lvert \widetilde{\overline{S}} \right\rvert^2 / \left\lvert \nabla \widetilde{\overline{\bm{u}}} \right\rvert^2$ confirms that the derived leading-order singularity can cause excessive values of the model constant of the DSM. The excessive values are relatively rare for forced isotropic turbulence. More than 99.0\% of the sample has a normalized model constant $C_{S} / \langle C_{S} \rangle$ in the range $[-25, 25]$, and the relative standard deviation of the model constant of the DSM is equal to 6.61. % 6.606711315502904.
To investigate the impact of the excessive values of the model constant of the local DSM on the sub-filter dissipation $\varepsilon_{m}$ in Eq.~(\ref{eq:modelsgsdiss}), the obtained range of values and the correlation with the exact sub-filter dissipation $\varepsilon_{\tau}$ in Eq.~(\ref{eq:exactsgsdiss}) are assessed. The sample has values of the normalized local sub-filter dissipation $\varepsilon_{m} / \langle \varepsilon_{m} \rangle$ in the range $[-969.3, 762.9]$, whereas the values of the normalized local exact sub-filter dissipation $\varepsilon_{\tau} / \langle \varepsilon_{\tau} \rangle$ are in the considerably narrower range $[-17.6, 51.1]$. The occurrence of modelled local sub-filter dissipation values considerably outside the range of the exact sub-filter dissipation values results in a relatively low correlation coefficient of 36.3\% between the local DSM and exact sub-filter dissipation.

The DNS results indicate that the derived singularity of the leading-order term of the model constant of the DSM is based on assumptions that are locally applicable also for turbulent flow. Therefore, the singularity is relevant for LES with the DSM. Because in a practical simulation, the calculation of the dynamic constant is based on the LES solution instead of the LES-filtered DNS solution, the above results do not directly imply a similarly severe singularity of the local dynamic model constant in practical simulations. However, in this paper, it is hypothesized that the singularity of the exact local model constant of the DSM can also result in large values of the local dynamic model constant in practical LES simulations. 

The above analysis focuses on the DSM, but a similar singularity can be derived for an eddy-viscosity model that locally approximates the exact sub-filter dissipation. An eddy-viscosity model assumes that the anisotropic part of the sub-filter model tensor is equal to
\begin{equation}
\label{eq:evmodel}
\tau_{i j} (\overline{\bm{u}}) - \frac{1}{3} \tau_{k k} (\overline{\bm{u}}) {\delta}_{i j} = - 2 \nu_{e} \overline{S}_{i j} \; ,
\end{equation}
where $\nu_{e}$ is the eddy viscosity \cite{pope-2000}. Setting the sub-filter dissipation provided by the eddy-viscosity model based on Eqs.~(\ref{eq:modelsgsdiss}) and (\ref{eq:evmodel}) equal to the series expansion of the exact sub-filter dissipation based on Eqs.~(\ref{eq:exactsgsdiss}) and (\ref{eq:clark}) gives a series expansion of the eddy viscosity
\begin{equation}
\label{eq:evtaylor}
\nu_{e} = - \frac{1}{12} \frac{ \overline{\Delta}_k^2 \left( \partial_k \overline{u}_i \right) \left(\partial_k \overline{u}_j \right) \overline{S}_{i j}}{ \left\lvert \overline{S} \right\rvert^2 } + {O} (\overline{\Delta}^2) \; .
\end{equation}
The leading-order term of this series expansion has a similar singularity as the dynamic model constant of the DSM. Thus, an eddy-viscosity model that locally approximates the exact sub-filter dissipation has a singular eddy viscosity.

\subsection{\label{sec:DSMdiscussion}Numerical stability of simulations with the local dynamic Smagorinsky model}
It is often reported that the dynamic constant of the DSM locally attains very large values if averaging is not applied, which causes numerical instability of simulations \cite{germano-1991,pope-2000}. The above analysis demonstrates that the exact dynamic model constant of the DSM attains very large values for flows for which the sub-filter dissipation of the Smagorinsky model in Eq.~(\ref{eq:smagorinsky_diss}) vanishes. However, the series expansion of the sub-filter dissipation of the DSM in Eq.~(\ref{eq:dsmdiss}) demonstrates that the model constant of the DSM locally approximates the exact sub-filter dissipation. This indicates that very large values of the model constant of the DSM are not necessarily unrealistic, but that they correct for coinciding non-zero exact sub-filter dissipation and vanishing dissipation of the Smagorinsky model. 

Accurate numerical simulation of the local cancellation of very large values of the model constant of the DSM and vanishing dissipation of the Smagorinsky model is not straightforward. This is because the rate-of-strain tensor in the Smagorinsky model in Eq.~(\ref{eq:smagorinsky_diss}) is effectively calculated based on a numerical discretization of the resolved rate-of-strain tensor, whereas the dynamic model constant is calculated based on a numerical discretization of the test filter. Because both terms are numerically approximated based on different discretizations and close to the grid cutoff, discretization errors can result in values of the dynamic model constant that are exceedingly large compared to the discretized dissipation. In this paper, it is hypothesized that very large values of the dynamic model constant can amplify discretization errors and that this is one of the causes of instability of simulations with the DSM if averaging is not applied. 

\section{\label{sec:LDMs}The dynamic gradient Smagorinsky model}
In this section, a local dynamic LES model with improved stability is derived by modifying the Smagorinsky model to remove the identified singularity of the model constant of the DSM.

The leading-order term of the series expansion of the model constant of the DSM in Eq.~(\ref{eq:cs_taylor_iso}) has a singularity primarily because the denominator of its leading-order term is a power of a contraction of the resolved rate-of-strain tensor which can vanish for a non-zero numerator. A straightforward approach to remove this singularity is to replace the resolved rate-of-strain tensor in the Smagorinsky model in Eq.~(\ref{eq:smagorinsky}) by the resolved velocity gradient tensor. This gives the sub-filter model tensor
\begin{equation}
\label{eq:gsmagorinsky}
\tau_{i j} (\overline{\bm{u}}) - \frac{1}{3} \tau_{k k} (\overline{\bm{u}}) {\delta}_{i j} = - 2 C_{\text{GS}} \overline{\Delta}^2  \left\lvert \nabla \overline{\bm{u}} \right\rvert \left( \partial_j \overline{u}_{i} \right) \; ,
\end{equation}
where $C_{\text{GS}}$ is the model constant and $\left\lvert \nabla \overline{\bm{u}} \right\rvert = \sqrt{2 \partial_i \overline{u}_{j} \partial_i \overline{u}_{j} }$. This model is called the gradient Smagorinsky model in this paper. The sub-filter dissipation provided by the gradient Smagorinsky model is
\begin{equation}
\label{eq:gsmagorinsky_diss}
\varepsilon_{m} = - \tau_{i j} (\overline{\bm{u}}) \left( \partial_j \overline{u}_i \right)  = C_{\text{GS}} \overline{\Delta}^2  \left\lvert \nabla \overline{\bm{u}} \right\rvert^3 \; .
\end{equation}
Unlike the Smagorinsky model, the gradient Smagorinsky model does not have a vanishing sub-filter dissipation if the resolved rate-of-strain tensor vanishes.

The static gradient Smagorinsky model is similar to the Smagorinsky model, but is modified to improve the stability of the corresponding local dynamic model. Similar to the Smagorinsky model, the gradient Smagorinsky model dissipates resolved kinetic energy for a positive model constant, primarily dissipates the kinetic energy of the smaller resolved scales, and can be derived based on dimensional analysis. The gradient Smagorinsky model only provides zero sub-filter dissipation if the resolved velocity gradient tensor vanishes, and thus provides non-zero sub-filter dissipation for more local flow conditions than the Smagorinsky model. It will be demonstrated that the additional sub-filter dissipation of the static gradient Smagorinsky model mitigates the singularity of the corresponding dynamic model.
The analyses presented in this paper could also be performed for other static LES models such as the gradient model \cite{vreman-1996,clark-1979}. However, other LES models are not considered in this research.

The sub-filter model tensor of the gradient Smagorinsky model is not intrinsically symmetric, unlike the exact sub-filter tensor and the sub-filter tensor of the Smagorinsky model. While this deviation can be considered to be fundamental, empirical and theoretical results indicate that the lack of intrinsic symmetry is not a material impediment for the purpose of deriving a dynamic LES model. The resolved rate-of-strain tensor used in the Smagorinsky model has a relatively low correlation with the exact physical sub-filter tensor \cite{clark-1979,mcmillan1979}. This indicates that functional LES models such as the Smagorinsky model have a considerable model error, which is expected to dominate the potential additional inaccuracy related to the lack of intrinsic symmetry of the gradient Smagorinsky model. Results obtained with the DGSM are found to be as good as results obtained with the DSM, which further indicates that the potential additional inaccuracy caused by the lack of symmetry of the gradient Smagorinsky model is sufficiently small to be addressed by application of the dynamic procedure.
In addition, while the gradient Smagorinsky model is not intrinsically symmetric, for incompressible flow its effect on the LES solution is similar to the effect of the slightly modified symmetric sub-filter model
\begin{equation}
\label{eq:gsmagorinskysymmetric}
\tau_{i j} (\overline{\bm{u}}) - \frac{1}{3} \tau_{k k} (\overline{\bm{u}}) {\delta}_{i j} = - 4 C_{\text{GS}} \overline{\Delta}^2  \left\lvert \nabla \overline{\bm{u}} \right\rvert  \overline{S}_{i j} \; .
\end{equation}
For this symmetric model the sub-filter term in Eq.~(\ref{eq:leseqs}) can be expressed as
\begin{equation}
\label{eq:gsmagorinskysymmetricequiv}
\partial_{j} \left( 4 C_{\text{GS}} \overline{\Delta}^2  \left\lvert \nabla \overline{\bm{u}} \right\rvert  \overline{S}_{i j} \right) = \partial_{j} \left( 2 C_{\text{GS}} \overline{\Delta}^2  \left\lvert \nabla \overline{\bm{u}} \right\rvert  \left( \partial_j \overline{u}_{i} \right) \right) + \partial_{j} \left( 2 C_{\text{GS}} \overline{\Delta}^2  \left\lvert \nabla \overline{\bm{u}} \right\rvert \right) \left( \partial_i \overline{u}_{j} \right) \; .
\end{equation}
The first term at the right-hand side is equal to the sub-filter term related to the gradient Smagorinsky model in Eq.~(\ref{eq:gsmagorinsky}). The second term reflects the difference caused by the lack of symmetry and is small if the eddy viscosity of the symmetric model $2 C_{\text{GS}} \overline{\Delta}^2  \left\lvert \nabla \overline{\bm{u}} \right\rvert$ has limited spatial variability. Thus, the effect of the gradient Smagorinsky model is expected to be similar to the effect of the symmetric model in Eq.~(\ref{eq:gsmagorinskysymmetric}). Results obtained in simulations (not shown) with the gradient Smagorinsky model in Eq.~(\ref{eq:gsmagorinsky}) and the symmetric model in Eq.~(\ref{eq:gsmagorinskysymmetric}) were found to be practically identical for the local dynamic model constant in Eq.~(\ref{eq:model_constant_dgsm}) derived below, which further demonstrates the similarity of the gradient Smagorinsky model to a symmetric model and the adequacy of the use of a sub-filter tensor that is not intrinsically symmetric. The symmetric model in Eq.~(\ref{eq:gsmagorinskysymmetric}) with the local dynamic model constant in Eq.~(\ref{eq:model_constant_dgsm}) is not further considered in this research because it is not a solution of the common contraction of the Germano identity.
In conclusion, the lack of intrinsic symmetry of the gradient Smagorinsky model is not considered to be a material impediment for deriving a dynamic LES model as the additional model error due to the lack of symmetry is relatively small, the deviation from a symmetric model is negligible for a sufficiently smooth eddy viscosity, and results obtained with the gradient Smagorinsky model are practically identical to results obtained with a symmetric model.

Application of the Germano identity to the gradient Smagorinsky model, making the same assumptions as in the derivation of the DSM, and application of the least-squares approach gives the dynamic model constant
\begin{equation}
\label{eq:model_constant_dgsm}
C_{\text{GS}} = \frac{L_{i j} M_{i j}'}{M_{k l}' M_{k l}'} \; ,
\end{equation}
where $L_{i j}$ is the Leonard tensor in Eq.~(\ref{eq:leonardtens}), and %WR: here reallywidetilde is used on second term
\begin{equation}
\label{eq:germano-identity_dgsmm}
M_{i j}' (\overline{\bm{u}}) = - 2 \widetilde{\overline{\Delta}}^2  \left\lvert \nabla \widetilde{\overline{\bm{u}}} \right\rvert \left( \partial_j \widetilde{\overline{u}}_{i} \right) + 2   \overline{\Delta}^2 \reallywidetilde{ \left\lvert \nabla \overline{\bm{u}} \right\rvert \left( \partial_j \overline{u}_{i} \right) } \; .
\end{equation}
Substitution of this dynamically determined model constant in the gradient Smagorinsky model in Eq.~(\ref{eq:gsmagorinsky}) gives the DGSM before any application of averaging or clipping.

To verify that the proposed gradient Smagorinsky model appropriately removes the singularity of the model constant of the DSM, the leading-order behavior of the model constant of the DGSM in Eq.~(\ref{eq:model_constant_dgsm}) is investigated. The series expansion of the tensor in Eq.~(\ref{eq:germano-identity_dgsmm}) is
\begin{equation}
\label{eq:germano-identity_dgsmm_taylor}
M_{i j}' (\overline{\bm{u}}) = - 2 \beta^2 \overline{\Delta}^2 \left\lvert \nabla \overline{\bm{u}} \right\rvert \partial_{j} \overline{u}_{i}  + {O} (\overline{\Delta}^4) \; .
\end{equation}
Substitution of this series expansion and the series expansion of the Leonard tensor in Eq.~(\ref{eq:leonard_taylor}) in the model constant of the DGSM in Eq.~(\ref{eq:model_constant_dgsm}) gives 
\begin{equation}
\label{eq:cgs_taylor}
C_{\text{GS}} = - \frac{1}{12} \frac{ \overline{\Delta}_k^2 \left( \partial_k \overline{u}_i \right) \left(\partial_k \overline{u}_j \right) \overline{S}_{i j}}{\overline{\Delta}^2 \left\lvert \nabla \overline{\bm{u}} \right\rvert^3} + {O} (\overline{\Delta}^2) \; .
\end{equation}
Substitution of this series expansion in the dissipation of the gradient Smagorinsky model Eq.~(\ref{eq:gsmagorinsky_diss}) and comparison with Eqs.~(\ref{eq:exactsgsdiss}) and (\ref{eq:clark}) demonstrates that the model constant of the DGSM locally approximates the exact sub-filter dissipation, similar to the model constant of the DSM. The numerator of the leading-order term of the series expansion in Eq.~(\ref{eq:cgs_taylor}) can be rewritten to the Frobenius inner product $\left( {\left( \nabla' \overline{\bm{u}} \right)}^{T} \left(\nabla' \overline{\bm{u}} \right) , \overline{S} \right)_F$, where ${\left( \nabla' \overline{\bm{u}} \right)}_{i j} = \overline{\Delta}_i \partial_i \overline{u}_j$ denotes the resolved velocity gradient tensor scaled by the LES filter width. By using the Cauchy-Schwarz inequality for the Frobenius inner product, the property $ \left\lVert (\nabla' \overline{\bm{u}})^{T} \nabla' \overline{\bm{u}} \right\rVert_{F} \leq \left\lVert \nabla' \overline{\bm{u}} \right\rVert_{F}^{2}$ of the Frobenius norm, the relation $\left\lVert A \right\rVert_{F} = \left\lvert A \right\rvert / \sqrt{2}$ of the Frobenius norm and the norm used in the Smagorinsky model, and the inequality $\left\lvert \nabla' \overline{\bm{u}} \right\rvert \le \max_{i} \left\{ \overline{\Delta}_{i} \right\} \left\lvert \nabla \overline{\bm{u}} \right\rvert$, an upper bound for the leading-order term of the series expansion of the model constant of the DGSM can be derived
\begin{eqnarray}
\label{eq:upper_bound_dgsm}
\left\lvert C_{\text{GS}} \right\rvert & \leq & \frac{1}{12} \frac{\left( {\left( \nabla' \overline{\bm{u}} \right)}^{T} \left(\nabla' \overline{\bm{u}} \right) , \overline{S} \right)_F}{ \overline{\Delta}^2 \left\lvert \nabla \overline{\bm{u}} \right\rvert^3 } + {O} (\overline{\Delta}^2) \leq \frac{1}{24} \frac{\left\lvert {\left( \nabla' \overline{\bm{u}} \right)}^{T} \left(\nabla' \overline{\bm{u}}\right) \right\rvert  \left\lvert \overline{S} \right\rvert}{ \overline{\Delta}^2 \left\lvert \nabla \overline{\bm{u}} \right\rvert^3 } + {O} (\overline{\Delta}^2) \nonumber\\
& \leq & \frac{1}{24 \sqrt{2}} \frac{\left\lvert \nabla' \overline{\bm{u}} \right\rvert^2  \left\lvert \overline{S} \right\rvert}{ \overline{\Delta}^2 \left\lvert \nabla \overline{\bm{u}} \right\rvert^3 } + {O} (\overline{\Delta}^2) \leq \frac{1}{24 \sqrt{2}} \max_{i} \left\{ \frac{\overline{\Delta}_i^2}{\overline{\Delta}^2 } \right\} + {O} (\overline{\Delta}^2) \; .
\end{eqnarray}
This demonstrates that the leading-order behavior of the model constant of the DGSM is not singular. A bounded dynamic model constant is expected to be a desirable property for the stability of the numerical simulation because it mitigates the occurrence of exceedingly large values. The above derivation of the upper bound does not utilize that for incompressible flow the trace of the resolved velocity gradient tensor is equal to zero. Therefore, the derived upper bound may not be tight for incompressible flow. 

To assess if the bounded leading-order term of the dynamic model constant of the DGSM removes the identified singularity for turbulent flow, the exact dynamic constant of the DGSM is calculated based on results of a DNS of forced isotropic turbulence \cite{perlman2007,Li2008}. The sample mean $\langle C_{\text{GS}} \rangle$ of the calculated dynamic model constant of the DGSM is equal to $7.41 \times 10^{-3}$, which is approximately half the sample mean of the dynamic models constant of the DSM. % obtained average value is 0.007406896909590189, the median is 0.0064837425
A scatter plot of the normalized dynamic model constant is shown in Fig.~\ref{fig:fig_dsm_dgsm_singularity}(b). Exceedingly large values of the normalized local dynamic model constant for vanishing $\left\lvert \widetilde{\overline{S}} \right\rvert^2 / \left\lvert \nabla \widetilde{\overline{\bm{u}}} \right\rvert^2$ are not observed for the DGSM. This indicates that the DGSM appropriately removes the identified singularity.
The relative standard deviation of the dynamic constant of the DGSM is equal to 1.41, which is considerably smaller than the relative standard deviation of the dynamic model constant of the DSM. % 1.4097651682071868
The obtained upper bound of the leading-order term of the dynamic constant of the DGSM in Eq.~(\ref{eq:upper_bound_dgsm}) is not satisfied for 2.2\% of the calculated exact values of the dynamic model constant. Thus, whereas assessment of the leading-order term suffices for analysis of the singularity, the higher-order terms are not negligible in general. % it is 2.22%
The sample has values of the normalized local sub-filter dissipation $\varepsilon_{m} / \langle \varepsilon_{m} \rangle$ in the range $[-72.6, 75.7]$, which is in better agreement with the range of exact sub-filter dissipation values than the range of values obtained for the local DSM. The correlation coefficient between the local DGSM sub-filter dissipation and the exact sub-filter dissipation is 68.1\%, which indicates moderate local correlation and is considerably higher than the correlation coefficient between the local DSM and the exact sub-filter dissipation.

For an anisotropic LES filter, the upper bound of the leading-order term of the dynamic model constant in Eq.~(\ref{eq:upper_bound_dgsm}) depends on the LES filter width. A dynamic model with an upper bound which is independent of the filter width can be obtained by setting the approximation of the LES filter width to $\overline{\Delta} = \max_i \left\{ \overline{\Delta}_i \right\}$ in the gradient Smagorinsky model in Eq.~(\ref{eq:gsmagorinsky}). This dynamic LES model could be appropriate for use in combination with an additional stabilization approach based on clipping of values of the dynamic model constant larger than a constant threshold. However, such a dynamic model is not further investigated in this paper.

The local model constant of the DGSM in Eq.~(\ref{eq:model_constant_dgsm}) has a bounded leading-order behavior, but can be negative. Therefore, it is expected that clipping negative values of the local model constant of the DGSM is still required to obtain a robust LES model. Therefore, the negative values of the dynamic model constant are clipped to zero
\begin{equation}
\label{eq:model_constant_dgsm1}
C_{\text{GS}} = \max \left\{ \frac{L_{i j} M_{i j}' }{M_{k l}' M_{k l}'} ,0 \right\} \; .
\end{equation}
In this paper, the DGSM with this model constant is called the local DGSM.

It will be demonstrated that simulations with the local DGSM have considerably better stability than simulations with the local DSM, and do not require averaging to prevent instability of simulations. This indicates that removal of the identified singularity of the model constant of the local DSM improves the stability of simulations. During this study, also an alternative local dynamic model was derived by means of normalization of the rate-of-strain tensor of the Smagorinsky model. Although the exact dynamic constant of this alternative model does not have a singular leading-order term, for anisotropic computational grids simulations with the model were found to be considerably less stable than simulations with the DGSM. This demonstrates that removal of the analytic singularity is not a sufficient condition for stability without averaging because dynamic LES models can be subject to other numerical or analytic instabilities. The rationale for using the gradient Smagorinsky model in Eq.~(\ref{eq:gsmagorinsky}) is that this LES model inaccurately provides excessive sub-filter dissipation for almost all numerical and analytic LES solutions. Because the dynamic procedure locally approximates the exact sub-filter dissipation, and the underlying static LES model already provides excessive sub-filter dissipation, it is expected that the obtained dynamic model constant vanishes for laminar flow and does not attain very large values which can result in amplification of discretization errors and numerical instability of simulations. Thus, the inaccuracy of the sub-filter dissipation provided by the gradient Smagorinsky model is what makes the model eligible for the application of the dynamic procedure without averaging. Conversely, application of the dynamic procedure to LES models that appropriately provide no sub-filter dissipation for laminar flow typically results in the instability of simulations even if averaging is applied \cite{bayatoda-2010}. It seems remarkable that stability of a dynamic model without averaging requires that the underlying static model provides excessive sub-filter dissipation, whereas this is generally considered to be an undesirable property of static LES models \cite{vreman-2004}.

The local DGSM has lower computational complexity than the averaged DSM and is not affected by the potential disadvantages of averaging. It will be demonstrated that satisfactory LES results can be obtained with the local DGSM. Thus, it is not necessary to apply averaging to the model constant of the local DGSM. Nevertheless, averaging the model constant of the DGSM is optional and can be applied if it is expected to increase the accuracy of simulation results. For example, if accurately capturing the global balance of production and dissipation of turbulent kinetic energy in a simulation is critical, the use of spatial averaging of the model constant of the DGSM can be considered. For a simulation of a statistically stationary flow, temporal averaging of the dynamic model constant can be considered.

\section{\label{sec:results}Results}
The above analytic derivation suggests that the local DGSM has improved stability compared to the local DSM and locally approximates the exact sub-filter dissipation. To verify that these desirable analytic properties also apply numerically, simulations of decaying homogeneous isotropic turbulence, a temporal mixing layer, and turbulent channel flow are performed. The simulations closely resemble simulations presented in previous assessments of LES models \cite{rozema-2015,lozanoduran-2019}.
The simulations are performed with second-order accurate discretizations of the incompressible Navier-Stokes equations for staggered rectangular computational grids \cite{verstappen-2003,bae-2018,bae-2019}. For the simulations of decaying homogeneous isotropic turbulence and the temporal mixing layer, time integration is performed using a one-leg method \cite{verstappen-2003} and for the simulations of turbulent channel flow the time integration is performed using a third-order Runge-Kutta method \cite{bae-2018,bae-2019}.

The local DGSM can be straightforwardly implemented by adjusting an existing implementation of the DSM. In this paper, the dynamic constant of the local DGSM in Eq.~(\ref{eq:model_constant_dgsm1}) and $\left\lvert \nabla \overline{\bm{u}} \right\rvert$ in the gradient Smagorinsky model in Eq.~(\ref{eq:gsmagorinsky}) are calculated at cell centers. The numerical solution is assumed to be equal to the LES solution $\overline{\bm{u}}$, and the LES filter is assumed to be a box filter. The test filter is approximated by a second-order accurate central discretization of a box filter with a filter width equal to twice the grid spacing $\widetilde{\Delta}_i = 2 {\Delta x}_{i}$, where ${\Delta x}_{i}$ is the grid spacing in the direction $x_{i}$. The Leonard tensor in Eq.~(\ref{eq:leonardtens}) and the tensor in Eq.~(\ref{eq:germano-identity_dgsmm}) are approximated at cell centers. The first term of the Leonard tensor is approximated by second-order central interpolation of the LES solution to cell centers and application of the discretized test filter. The second term of the Leonard tensor is approximated by application of the discretized test filter to the staggered LES solution, and second-order interpolation to cell centers. The tensor in Eq.~(\ref{eq:germano-identity_dgsmm}) is approximated by second-order central discretization of the velocity gradient at cell centers, and application of the discretized test filter at cell centers. Based on these approximations, the constant of the local DGSM in Eq.~(\ref{eq:model_constant_dgsm1}) can be calculated at cell centers. It is recommended to mitigate undesirable propagation of floating-point rounding errors by adding a very small number to the denominator of the dynamic model constant. The filter width approximation used in the LES model is set equal to $\overline{\Delta}^3 = {\Delta x}_{1} {\Delta x}_{2} {\Delta x}_{3}$, and the filter width of subsequent application of the LES and test filters is assumed to satisfy $\widetilde{\overline{\Delta}} = \alpha \overline{\Delta}$, for some parameter $\alpha$ \cite{germano-1991}. The parameter $\alpha$ is the only parameter of the dynamic models and reflects the relation of the implicit LES filter of the used numerical method and the discretized test filter. Accurate results with the DSM have previously been obtained with the used simulation methods for $\alpha = 2$ \cite{rozema-2015}. Therefore, this value is also used for the local DGSM.

Simulations with LES models that locally approximate the exact sub-filter dissipation typically require a smaller time step size for numerical stability of the time integration than simulations with models that globally provide the appropriate level of sub-filter dissipation. Therefore, it is expected that simulations with the DGSM require a smaller time step size for stability than simulations with the averaged DSM. The simulations with the local DGSM in this paper are performed at time step sizes similar to time step sizes that were previously used for simulations with the AMD model \cite{rozema-2015,lozanoduran-2019}. 

\subsection{\label{sec:resultsHIT}Decaying homogeneous isotropic turbulence}
To assess the applicability of the local dynamic models for homogeneous isotropic turbulence, simulations of the experiment by Comte-Bellot and Corrsin are performed \cite{comte-bellot-1971}. In this experiment, turbulence is generated by a grid with mesh size $M = 5.08 \textnormal{ cm}$ in a flow of mean velocity $U_0 = 1000 \textnormal{ cm}/\textnormal{s}$. Energy spectra are recorded at three locations $42 M$, $98 M$, and $171 M$ downstream of the grid. The simulations consider the flow inside a cube of length $11 M$ which moves along with the mean flow and is located at the grid at $t=0 \textnormal{ s}$. Therefore, the energy spectrum of the simulated turbulence is expected to be identical to the LES-filtered measured energy spectra at $t = 42 M / U_0$, $t = 98 M / U_0$, and $t = 171 M / U_0$. All quantities are non-dimensionalized by the length of the cube $L_{\textnormal{ref}} = 11 M = 55.88 \textnormal{ cm}$ and a reference velocity $u_{\textnormal{ref}} = 27.19 \textnormal{ cm} / \textnormal{s}$ which satisfies $u_{\textnormal{ref}}^2 = 3 \langle u_1'^2 \rangle / 2$ at the first measurement station in the experiment, where the brackets denote averaging in all the spatial directions and the prime denotes fluctuations with respect to this average. The computational grid is isotropic with 64 cells in each direction. The time step size is set to ${\Delta t} = 1.59 \times 10^{-3} L_{\textnormal{ref}} / u_{\textnormal{ref}}$. The Reynolds number is $\textnormal{Re} = u_{\textnormal{ref}} L_{\textnormal{ref}} / \nu_{\textnormal{ref}} = 10129$. The initial condition is generated following the procedure described in the paper that introduced the AMD model \cite{rozema-2015}. For the dynamic model constant of the averaged DSM, averaging is applied in all the spatial directions. The total resolved kinetic energy $E = \overline{u}_i \overline{u}_i / 2$ and the resolved kinetic energy spectra $E(\kappa)$ as a function of the wave number magnitude $\kappa$ obtained with the local DSM, averaged DSM, and local DGSM are calculated and compared with box-filtered experimental measurements at $t = 98 M / U_0$ and $t = 171 M / U_0$. To further investigate potential differences between the DSM and DGSM, the resolved vorticity $\overline{\boldsymbol{\omega}} = \nabla \times \overline{\boldsymbol{u}}$, the total resolved enstrophy $\left\lVert \overline{\Omega} \right\rVert_{F}^2 = \overline{\Omega}_{i j} \overline{\Omega}_{i j} = \overline{\omega}_i \overline{\omega}_i / 2$, and the resolved enstrophy spectra $\left\lVert \Omega \right\rVert _{F}^{2} (\kappa)$ are also calculated.

\begin{figure*} % WR: in the APS template often the option [b] is used
\includegraphics{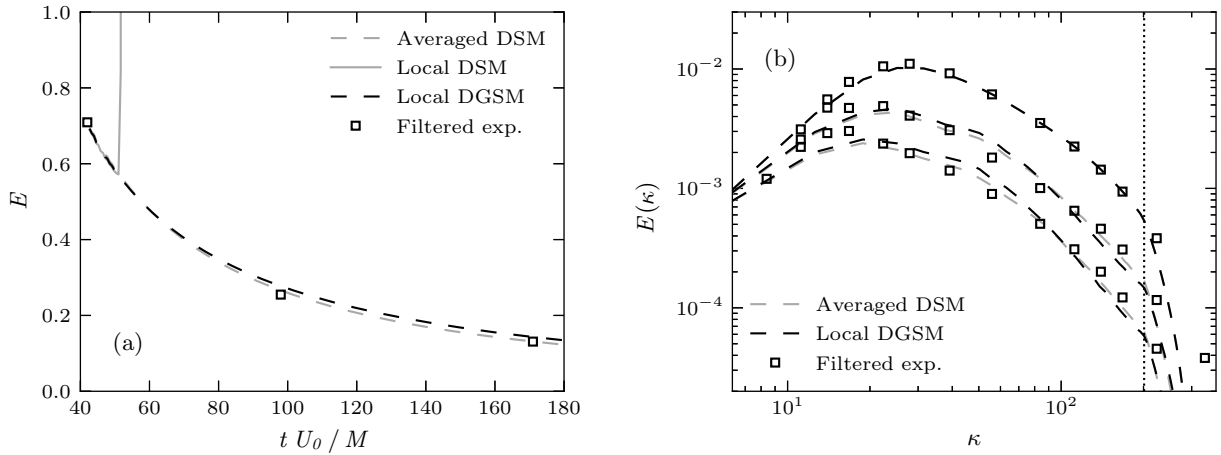}
\caption{\label{fig:cbc} The total resolved kinetic energy (a) and the resolved energy spectra at $t = 42 M / U_0$, $t = 98 M / U_0$, and $t = 171 M / U_0$ (b) obtained with the DSM and DGSM, and the box-filtered experimental measurements. The vertical dotted line represents the largest wave number that can be represented on the computational grid.} % add about filtere results
\end{figure*}

The total resolved kinetic energy and resolved kinetic energy spectra obtained with the DSM and DGSM are shown in Fig.~\ref{fig:cbc}. 
The simulation with the local DSM is unstable and gives unrealistic results from approximately $t = 50 M / U_0$. The simulation with the local DGSM is stable. The results obtained with the local DGSM and averaged DSM are almost identical, and accurately agree with box-filtered experimental measurements. The total resolved enstrophy and resolved enstrophy spectra obtained with the averaged DSM and local DGSM are also almost identical, and the obtained resolved enstrophy spectra accurately agree with box-filtered experimental measurements for wave numbers that can be represented on the computational grid (not shown).

To assess the occurrence of large values of the dynamic model constant of the local DSM and local DGSM, time series of the maximum value of the dynamic model constants normalized by the spatial average of the dynamic model constants are investigated. For the local DSM the temporal average of $\max \{ C_{S} \} / \langle C_{S} \rangle$ before the simulation becomes unstable is equal to 127.5. For the local DGSM the temporal average of $\max \{ C_{\text{GS}} \} / \langle C_{\text{GS}} \rangle$ is equal to 47.2. Thus, the normalized maximum value of the dynamic model constant is considerably larger for the local DSM than for the local DGSM. This indicates that the local DGSM mitigates the occurrence of large relative values of the dynamic model constant in practical LES.

In the simulations of decaying homogeneous isotropic turbulence, results obtained with the local DGSM are similar to results obtained with the averaged DSM and accurately approximate the experimental measurements. Whereas the simulation with the local DSM is unstable, the local DGSM does not require averaging to prevent the instability of the simulation. This suggests that the local DGSM is a dynamic model which does not require averaging for stability or accuracy of LES results.

\subsection{\label{sec:resultsMixingLayer}Temporal mixing layer}
To assess the applicability of the local dynamic models to transitional flow, simulations of a temporal mixing layer are performed. A temporal mixing layer consists of two streams with opposite flow velocities. A Kelvin-Helmholtz instability originates at the interface of the two streams and causes the transition to turbulence of the mixing layer. A temporal mixing layer is a simplified model of a spatial shear layer. A turbulent temporal mixing layer is expected to be self-similar for some time \cite{rogers-1994}. 

The simulations of an incompressible temporal mixing layer performed in this paper are similar to previously performed simulations of a weakly compressible mixing layer at a high Reynolds number \cite{vreman-2004}. The $x_{1}$-axis is aligned with the stream-wise direction, the $x_{2}$-axis is aligned with the direction normal to the mixing layer, and the $x_{3}$-axis is aligned with the span-wise direction. All the quantities are non-dimensionalized by half the initial vorticity thickness of the mixing layer and the far-field stream-wise velocity. The initial dimensionless LES-filtered velocity field is based on a hyperbolic tangent
\begin{equation}
\label{eq:initml}
\overline{u}_{1} = \tanh \left( x_{2} \right) \; , \quad \overline{u}_{2} = \overline{u}_{3} = 0 \; ,
\end{equation}
and random perturbations with an amplitude of $0.05 e^{-x_{2}^{2} /4}$ are added near the center plane of the mixing layer to trigger the transition to turbulence. The details of the transition and growth of the temporal mixing layer are observed to be sensitive to the realization of the added random perturbations. Therefore, the same perturbations have been used for the simulations with the different LES models. 

The Reynolds number based on half the initial vorticity thickness is $\textnormal{Re} = 100 000$. The computational domain spans $90$ times half the initial vorticity thickness of the mixing layer in each direction. The simulations are performed on isotropic and anisotropic rectangular computational grids with constant grid spacing in each direction. The computational grids have 128 cells in the direction normal to the mixing layer and 128, 32, or 16 grid cells in the stream-wise and span-wise directions. Thus the cells of the computational grids have aspect ratios ${\Delta x}_1 / {\Delta x}_2 = {\Delta x}_3 / {\Delta x}_2$ equal to 1, 4, and 8. The non-dimensionalized time step size $\Delta t$ is set equal to $1.0 \times {10}^{-1}$, $6.0 \times {10}^{-2}$, and $3.0 \times {10}^{-2}$ for the simulations on grids with aspect ratios of 1, 4, and 8, respectively. The boundary conditions in the stream-wise and span-wise directions are periodic, and in the direction normal to the mixing layer a free-slip boundary condition is imposed. For the dynamic model constant of the averaged DSM, averaging is applied in the stream-wise and span-wise directions.

Because the details of the transition and growth of the temporal mixing layer are sensitive to the realization of the added random perturbations and representative DNS results are not available, the assessment of the LES models for the temporal mixing layer is primarily qualitative. Results obtained with the considered LES models are compared, with specific attention for the moment of transition to turbulence, the self-similarity of LES solutions, and undesirable accumulation of resolved kinetic energy at the smallest resolved scales in the LES solution. To investigate the growth of the mixing layer, the momentum thickness of the mixing layer
\begin{equation}
\label{eq:momthickness}
\theta = \frac{1}{4} \int_{-\infty}^{\infty} \left( 1- \langle \overline{u}_{1} \rangle \right) \left( 1+ \langle \overline{u}_{1} \rangle \right) \textnormal{d} x_2
\end{equation}
is calculated, where the brackets denote averaging in the stream-wise and span-wise directions. The results obtained with the local DSM, averaged DSM, and local DGSM are compared with results obtained with the AMD model \cite{rozema-2015}. The AMD model is a static LES model which provides no sub-filter dissipation for laminar flow.

\begin{figure*}
\includegraphics{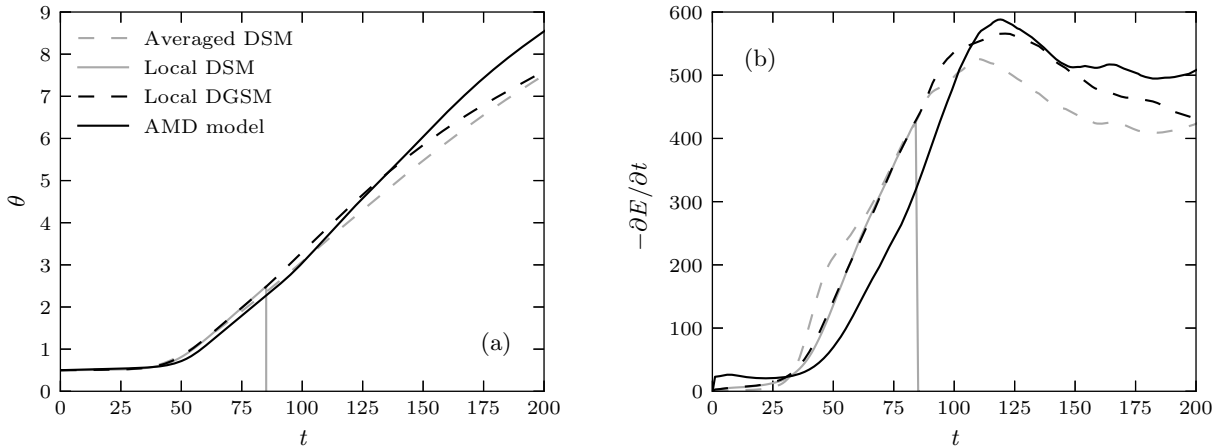}
\caption{\label{fig:ml_growth} The momentum thickness of the mixing layer (a) and the total resolved kinetic energy dissipation rate (b) obtained with the local DSM, averaged DSM, local DGSM, and AMD model.}
\end{figure*}

First, the results of the simulations on the isotropic computational grid are presented. The momentum thickness of the mixing layer and the resolved kinetic energy dissipation rate obtained with the local DSM, averaged DSM, local DGSM, and AMD model are shown in Fig.~\ref{fig:ml_growth}.
The simulation with the local DSM is unstable and gives unrealistic results from approximately $t = 85$. The simulation with the local DGSM is stable, and the obtained results closely agree with results obtained with the averaged DSM. All the considered LES models predict an approximately linear increase of the momentum thickness of the mixing layer between $t = 60$ and $t=150$. This suggests that the mixing layer is self-similar for some time after the transition to turbulence. The local DGSM provides almost no dissipation of resolved kinetic energy before $t = 25$, which suggests that the local DGSM appropriately provides no sub-filter dissipation for laminar flow. The local DGSM predicts transition at approximately the same moment as the averaged DSM, and slightly before the transition moment predicted by the AMD model. This indicates that averaging is not required to appropriately predict transition to turbulence with dynamic models.

\begin{figure*}
\includegraphics{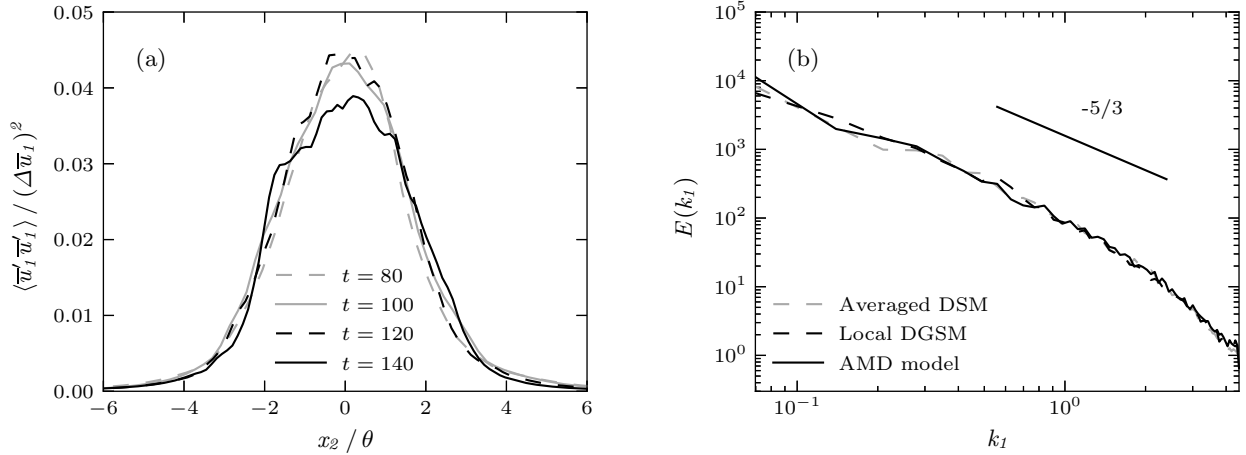}
\caption{\label{fig:ml_spec_selfsimdgsm} The stream-wise velocity fluctuations obtained with the local DGSM at different times after transition to turbulence (a) and the stream-wise kinetic energy spectra at the center plane of the temporal mixing layer at $t=140$ obtained with the averaged DSM, local DGSM, and the AMD model (b).}
\end{figure*}

To further assess the self-similarity of the solution obtained with the local DGSM, the stream-wise velocity fluctuations are investigated. The stream-wise velocity fluctuations are defined as $\langle \overline{u}_1' \overline{u}_1' \rangle$, where the brackets denote averaging in the stream-wise and span-wise directions and the prime denotes fluctuations with respect to this average. Plots of the normalized stream-wise velocity fluctuations obtained with the local DGSM against the normalized coordinate normal to the mixing layer are shown in Fig.~\ref{fig:ml_spec_selfsimdgsm}(a).
Plots of the stream-wise velocity fluctuations obtained with the local DGSM model at different times almost fully collapse, which demonstrates that the model appropriately captures the self-similar nature of the turbulent mixing layer.

To assess if the proposed local DGSM provides sufficient sub-filter dissipation to prevent the accumulation of resolved kinetic energy at the smallest resolved scales, spectra of the resolved kinetic energy are investigated. The spectra of the resolved kinetic energy in the stream-wise direction at the center plane of the mixing layer at $t = 140$ are shown in Fig.~\ref{fig:ml_spec_selfsimdgsm}(b). The kinetic energy spectra obtained with all the considered LES models appropriately satisfy the desired $E(k_{1}) \propto k_{1}^{-5/3}$ decay rate, where $k_{1}$ is the stream-wise wave number. Pile-up of resolved kinetic energy is not observed, which indicates that the local DGSM provides sufficient sub-filter dissipation.

\begin{figure*}
\includegraphics{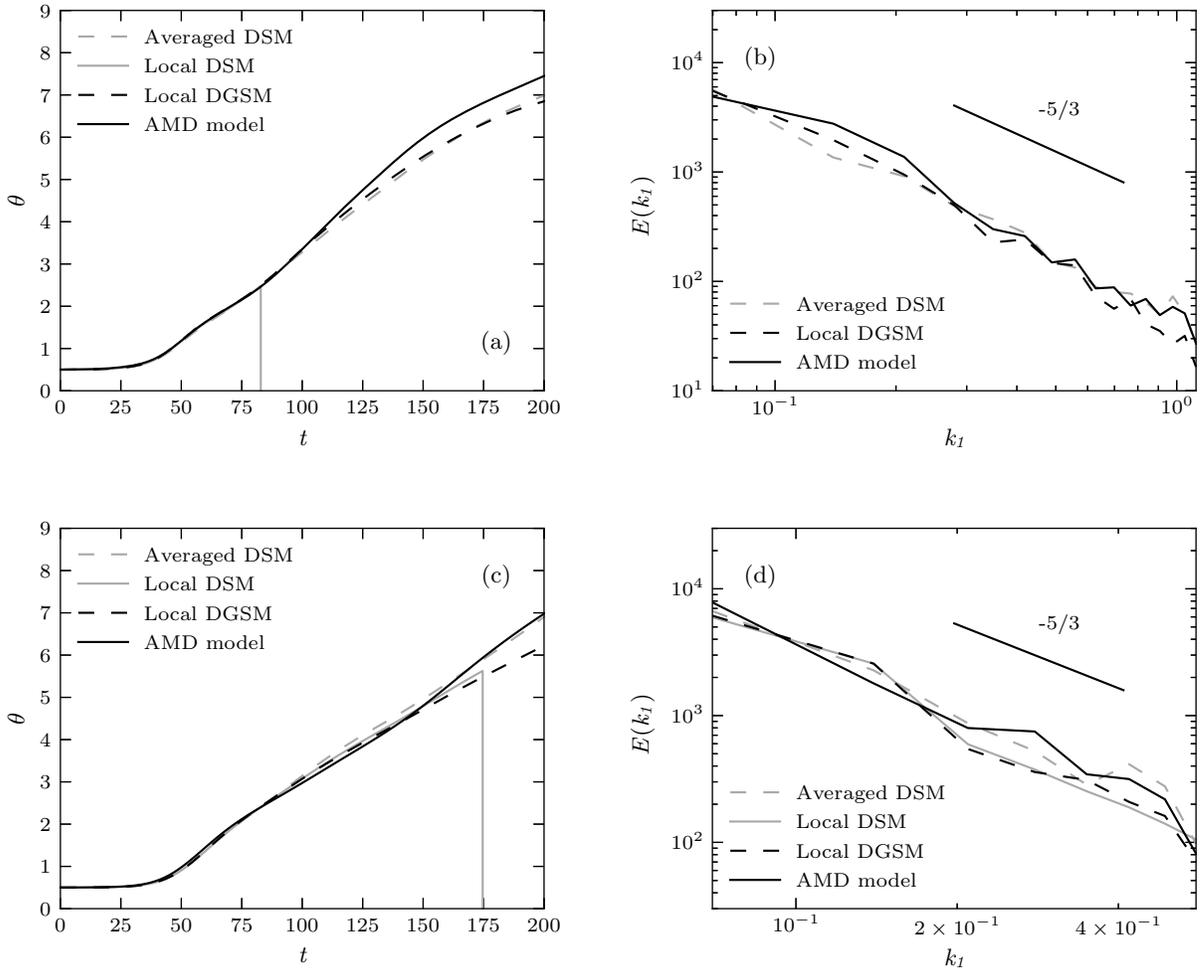}
\caption{\label{fig:ml_ar4_ar8} The momentum thickness of the mixing layer and the stream-wise kinetic energy spectra at the center plane of the temporal mixing layer at $t=140$ obtained in simulations of the temporal mixing layer on the computational grids with aspect ratios of 4 (a) and (b), and 8 (c) and (d).}
\end{figure*}

Results of the simulations on the anisotropic computational grids with aspect ratios of 4 and 8 are shown in Fig.~\ref{fig:ml_ar4_ar8}.
The results obtained in the simulations on the anisotropic grids are generally similar to results obtained in the simulations on the isotropic grid. The simulations with the local DSM are unstable for both the anisotropic grids. The simulations with the local DGSM are stable, and the obtained results closely agree with results obtained with the averaged DSM for both the anisotropic grids. The local DGSM predicts an approximately linear increase of the momentum thickness of the mixing layer between $t = 60$ and $t=150$, in good agreement with the other considered LES models. The kinetic energy spectra obtained with all the considered LES models appropriately satisfy the desired decay rate, which indicates that the local DGSM provides sufficient sub-filter dissipation also on anisotropic grids.

To further investigate potential differences between the local DGSM and the averaged DSM, the total resolved enstrophy and resolved streamwise enstrophy spectra have also been compared for the simulations of the mixing layer on isotropic and anisotropic computational grids (not shown). The local DGSM and averaged DSM are found to provide similar results for the enstrophy.

Whereas the performed simulations of the temporal mixing layer with the local DSM are unstable, simulations with the local DGSM are stable, and results obtained with the local DGSM closely agree with results obtained with the averaged DSM for isotropic and anisotropic computational grids. This suggests that the local DGSM is an appropriate local dynamic LES model for the simulation of transitional flow on both isotropic and anisotropic computational grids.

\subsection{\label{sec:resultsChannel}Turbulent channel flow} 
To assess the applicability of the local dynamic models to wall-bounded flow, simulations of turbulent channel flow are performed. To assess the performance of the models for wall-resolved LES the mean flow velocity profile, turbulent fluctuations, and turbulent kinetic energy budgets obtained in simulations of channel flow at a low Reynolds number are compared with results from a DNS. To assess the performance of the models for a wall-modeled LES, the scaling of errors of the mean flow velocity in the outer region of wall-bounded turbulence is assessed for simulations of channel flow at moderate Reynolds numbers. Convergence of the error in the outer region of wall-bounded turbulence is considered to be important because for wall-modeled LES the LES model is primarily expected to accurately predict the sub-filter dissipation in the outer region, rather than to predict the wall-shear stress \cite{lozanoduran-2019}.

For the considered turbulent channel flows, the $x_{1}$-axis is aligned with the stream-wise direction, the $x_{2}$-axis is aligned with the wall-normal direction, and the $x_{3}$-axis is aligned with the span-wise direction. Brackets denote averaging in the stream-wise, span-wise, and temporal directions, and a prime denotes fluctuations with respect to this average. 

For the dynamic model constant of the averaged DSM, averaging is applied in the stream-wise and span-wise directions. Simulations of turbulent channel flow with the averaged DSM sometimes apply the LES and test filter only in the stream-wise and span-wise directions \cite{germano-1991}. To assess the performance of the proposed LES models for general wall-bounded flows, in this paper the LES and test filter are also applied in the wall-normal direction. In the interior of the channel, the test filter can be applied straightforwardly in the wall-normal direction. For grid cells at the wall, the ratio $\alpha$ of the LES and test filter widths are kept constant, and the discretization of the test filter depends on flow quantities in dummy cells beyond the wall boundary. The boundary condition of the LES solution is applied to set the flow quantities in the dummy cells. Some sensitivity of the LES results is observed to the discretization of the test filter at the wall. For wall-resolved LES the wall-normal grid spacing is relatively small at the wall, and therefore the sensitivity to the discretization is small for wall-resolved LES.

\subsubsection{\label{sec:resultsChannelWR}Wall-resolved LES at a low Reynolds number}
To assess the applicability of the local DGSM for wall-resolved LES, simulations of turbulent channel flow at a friction Reynolds number of $\textnormal{Re}_{\tau} \approx 590$ are performed. The dimensions of the considered channel are $2 \pi \delta \times 2 \delta \times \pi \delta$, where $\delta$ is the channel half-height. The channel flow is driven by a constant pressure gradient. The initial condition is set to a Poiseuille flow and divergence-free perturbations of small amplitudes are added to trigger the transition to turbulence. After the flow has transitioned to turbulence, flow statistics are recorded and compared with results of a DNS \cite{moser-1999}. Results of the channel flow simulations are normalised by viscous scales based on the kinematic viscosity $\nu$ and the friction velocity $u_\tau = \sqrt{\tau_w/\rho}$, where $\rho$ is the density of the fluid and $\tau_w$ is the wall shear stress \cite{pope-2000}. Quantities normalized by viscous scales are denoted by a plus sign. For example, the wall-normal coordinate measured in viscous length scales is denoted $x_{2}^{+} = x_{2} u_\tau  / \nu$. 

The computational grid is relatively coarse with $64$ cells in each direction. The grid stretches towards the wall in the wall-normal direction according to a hyperbolic tangent distribution, and the grid spacing is uniform in the stream-wise and span-wise directions \cite{bae-2018,bae-2019}. The height of the first grid cell at the wall measured in viscous length scales is ${\Delta x}_{2}^{+} = 3.9$. The time step size of the simulations is set dynamically based on the Courant–Friedrichs–Lewy condition for numerical stability $\Delta t \le \min_i \left\{ {\Delta x}_{i} \right\} /\lVert \overline{\bm{u}} \rVert$, where $\lVert \overline{\bm{u}} \rVert$ is the resolved velocity magnitude. The boundary conditions in the stream-wise and span-wise directions are periodic, and at the walls of the channel, a no-slip boundary condition is imposed.

\begin{figure*}
\includegraphics{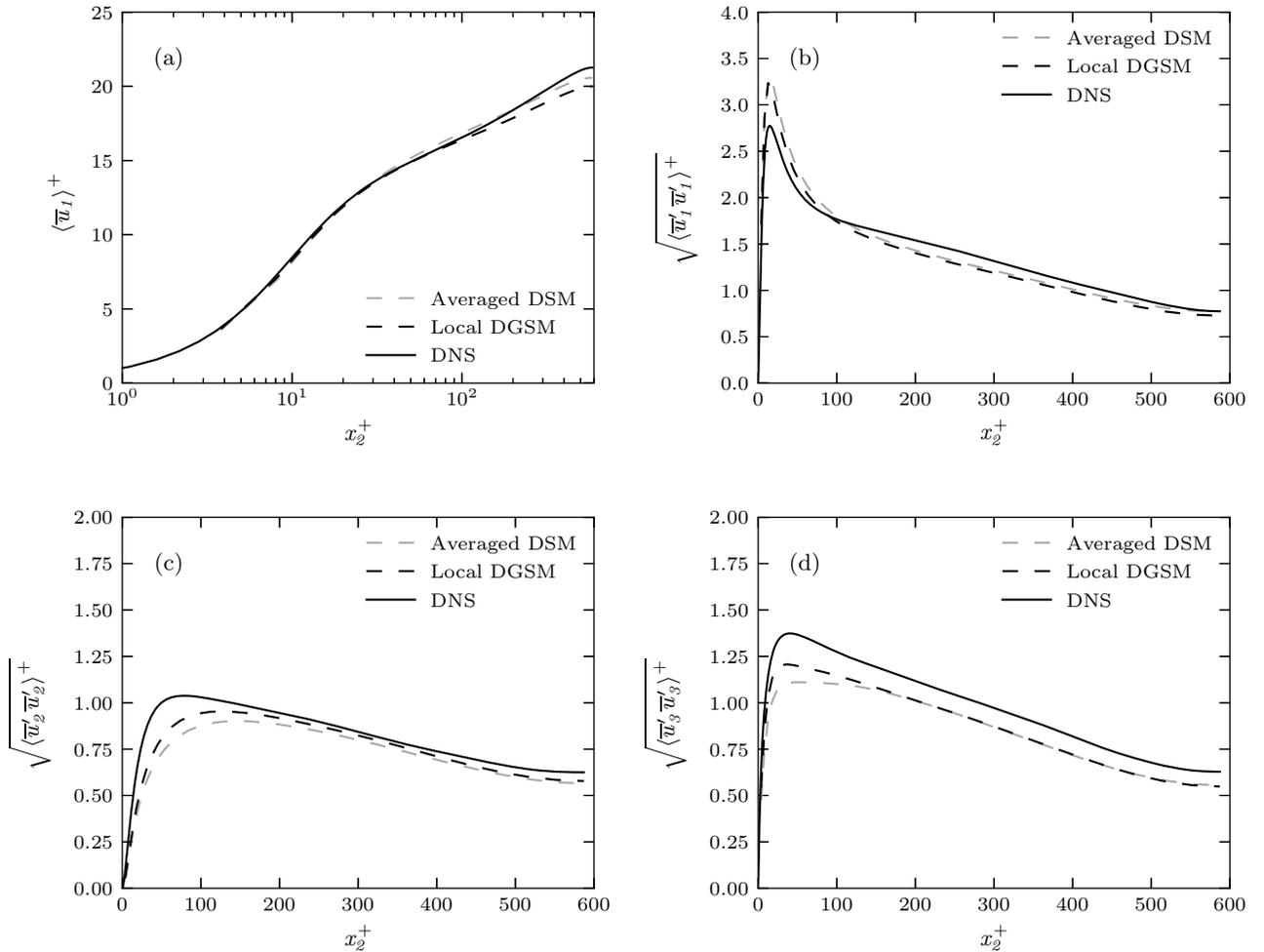}
\caption{\label{fig:chan_wall_resolved} The normalized mean flow velocity (a), stream-wise turbulent fluctuations (b), wall-normal turbulent fluctuations (c), and span-wise turbulent fluctuations (d) obtained with the averaged DSM and local DGSM.}
\end{figure*}

The channel flow simulation with the local DGSM is stable, whereas the simulation with the local DSM is unstable. The values of the bulk velocity $U_{b} = \int_0^\delta \langle u_1 \rangle \mathrm{d}x_2/\delta$ obtained with the averaged DSM and local DGSM are listed in Table~\ref{tab:table_channel_retau}. The mean stream-wise flow velocity and the turbulent fluctuations are shown as a function of the wall-normal coordinate in Fig.~\ref{fig:chan_wall_resolved}.
\begin{table*}
\caption{\label{tab:table_channel_retau} The bulk velocity $U_{b}^{+}$ obtained in simulations of turbulent channel flow and the relative error compared to the bulk velocity obtained in a DNS.}
\begin{ruledtabular}
\begin{tabular}{lccc}
& Averaged DSM & Local DGSM & DNS \\
\hline
$U_{b}^{+}$ & 18.40 & 17.92 &  18.65 \\
Relative error & -1.36\% & -3.95\% & 
\end{tabular}
\end{ruledtabular}
\end{table*}
The bulk flow velocity, mean flow velocity, and turbulent fluctuations obtained with the averaged DSM and local DGSM are in good agreement with DNS results. The averaged DSM predicts the bulk flow velocity slightly more accurately than the local DGSM, whereas the local DGSM predicts the turbulent fluctuations close to the wall slightly more accurately.

To further assess the performance of the proposed LES models, the resolved turbulent kinetic energy budgets are investigated. The resolved turbulent kinetic energy is
\begin{equation}
\label{eq:tkedef}
\overline{e} = \frac{1}{2} \left( {\overline{u}'}^2 + {\overline{v}'}^2+ {\overline{w}'}^2\right) \; .
\end{equation}
The resolved turbulent kinetic energy budget equation for LES is
\begin{equation}
    0 = \mathcal{P} + \varepsilon + \varepsilon_{m} + T_\mathrm{turb}+ T_p + T_\nu \; ,
\end{equation}
where 
\begin{eqnarray}
\label{eq:turb_tke_defs}
& \mathcal{P} = -\left\langle{\overline{u}_1'\overline{u}_{2}'}\right\rangle \frac{\mathrm{d} \left\langle \overline{u}_1 \right\rangle}{\mathrm{d} x_2} \; ,\quad \varepsilon = - \nu\left\langle{\left(\frac{\partial \overline{u}_i'}{\partial x_j}\right) \left(\frac{\partial \overline{u}_i'}{\partial x_j}\right) }\right\rangle \; , \quad \varepsilon_{m} = - \left\langle{\overline{u}_i'\frac{\partial\tau_{ij}'}{\partial x_j}}\right\rangle \; , & \nonumber\\
& T_\mathrm{turb} = - \frac{\mathrm{d}\left\langle{\overline{e}\,\overline{v}'}\right\rangle}{\mathrm{d} x_2} \; , \quad T_p = - \frac{1}{\rho}\frac{\mathrm{d} \left\langle{\overline{p}'\overline{u_2}'}\right\rangle}{\mathrm{d} x_2} \; , \quad T_{\nu} = \nu\frac{\mathrm{d}^2 \left\langle{\overline{e}} \right\rangle}{\mathrm{d} x_2^2} &
\end{eqnarray}
are the resolved turbulent kinetic energy production, viscous dissipation, sub-filter dissipation provided by the LES model, turbulent transport, pressure diffusion, and viscous transport of resolved turbulent kinetic energy, respectively \cite{pope-2000,cho-2018}. Because averaging is applied in all the directions except for the wall-normal direction, the terms in the budget equation depend only on the wall-normal coordinate. 

\begin{figure}
\includegraphics{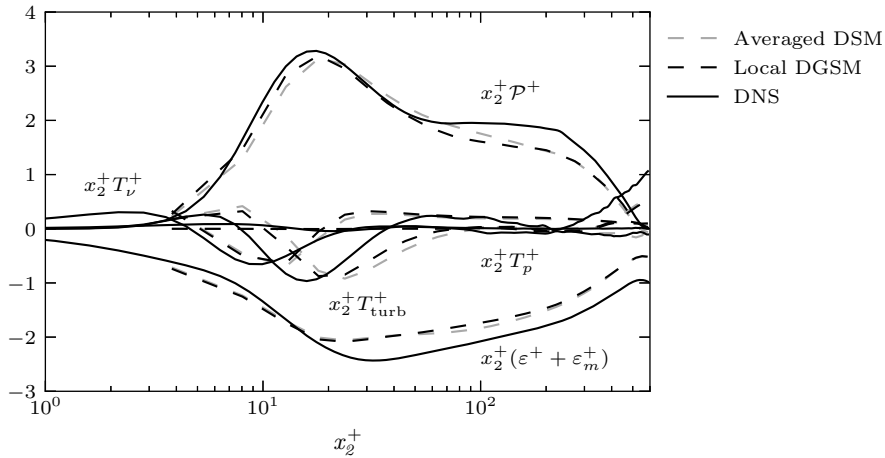}
\caption{\label{fig:chan590_energy_budgets} The terms of the turbulent kinetic energy budget obtained with the averaged DSM and local DGSM for the channel flow at $\textnormal{Re}_\tau \approx 590$.}
\end{figure}

The turbulent kinetic energy budget terms obtained in simulations with the averaged DSM and local DGSM are shown in Fig.~\ref{fig:chan590_energy_budgets}. Because all the considered LES models are dissipative, the viscous dissipation obtained in the DNS is compared with the sum of the viscous dissipation and sub-filter dissipation provided by the LES model. This comparison assumes that the unresolved sub-filter dissipation is small compared to the resolved viscous dissipation.
The terms of the energy budget equation obtained with the local DGSM are practically identical to the terms obtained with the averaged DSM. Results obtained with the considered LES models reasonably agree with DNS results. 

The simulation with the local DGSM is stable for the considered wall-resolved LES of turbulent channel flow. The results obtained with the local DGSM are similar to results obtained with the averaged DSM. The averaged DSM provides slightly more accurate predictions of the bulk flow velocity, whereas the local DGSM provides slightly more accurate predictions of the turbulent fluctuations close to the wall.

\subsubsection{\label{sec:resultsChannelWM}Wall-modeled LES of the outer region at a moderate Reynolds number}
To assess the performance of the local DGSM in the outer region of wall-bounded turbulence, simulations of turbulent channel flow with an exact wall model are performed at friction Reynolds numbers of $\textnormal{Re}_\tau \approx 4200$ and $\textnormal{Re}_\tau \approx 8000$. The simulations closely resemble simulations presented in the paper that introduced a methodology to assess the scaling of LES errors in the outer region of wall-bounded turbulence \cite{lozanoduran-2019}. The simulations are generally similar to the wall-resolved simulations of channel flow for the interior of the channel. However, a Neumann boundary condition
\begin{equation}
\label{eq:bcouterassessment}
\frac{\partial \overline{u}_1}{\partial n} = \frac{\tau_{w} - \tau_{1 2} }{\nu} \;
\end{equation}
is applied at the wall of the channel, where $n$ is the wall-normal direction pointing towards the interior of the channel domain and $\tau_{12}$ is the shear stress provided by the LES model. The channel flow is driven by a constant bulk velocity obtained in a DNS, and the desired near-wall behavior of the LES solution is imposed by setting the wall shear stress $\tau_{w}$ equal to the averaged wall shear stress obtained in a DNS \cite{lozanoduran-2014}. The applied Neumann boundary condition can be considered to be an exact averaged wall model.

Simulations are performed for three grid resolutions to investigate the convergence of the error of the mean velocity profile in the outer region predicted by LES models. The computational grids are approximately isotropic ${\Delta x}_1 \approx {\Delta x}_2 \approx {\Delta x}_3$ and the three considered grids have grid spacings of ${\Delta x}_2 = \delta/5$, ${\Delta x}_2 = \delta/10$, and ${\Delta x}_2 = \delta/20$. The relative error in the mean stream-wise velocity profile in the outer region is defined as
\begin{equation}
    \mathcal{E} = \sqrt{ \frac{\int_{\delta/5}^{\delta} \left(\left\langle \overline{u}_1 \right\rangle - \left\langle u_1 \right\rangle^{\mathrm{DNS}} \right) ^2 \mathrm{d} x_{2}} {\int_{\delta/5}^{\delta} \left(\left\langle u_1 \right\rangle^{\mathrm{DNS}}\right)^2 \mathrm{d} x_{2}} } \; ,
    \label{eq:mean_error}
\end{equation}
where $\left\langle \overline{u}_1 \right\rangle$ is the averaged stream-wise flow velocity in the LES and $\left\langle u_1 \right\rangle^{\mathrm{DNS}}$ is the mean stream-wise velocity obtained in a DNS \cite{lozanoduran-2014,yamamoto-2018}. This error measure does not take into account errors in the region $x_{2} < \delta / 5$ close to the wall. 

Investigation of the convergence of the error measure in Eq.~(\ref{eq:mean_error}) assesses the ability of LES models to predict the mean flow velocity in the outer region of wall-bounded turbulence if an exact wall model is applied. An assumption of the assessment is that the flow in the outer region depends primarily on the average wall shear stress imposed by the Neumann boundary condition, and is largely independent of the details of the turbulence close to the wall. For the considered turbulent channel flow this assumption is supported by laboratory and numerical experiments in which the near-wall flow was modified or rough-wall boundary conditions were introduced without affecting the mean profile in the outer region \cite{nikuradse-1933,perry-1977,lee-2013,flores-2006,chung-2014,lozanoduran-2019a}.

\begin{figure}
\includegraphics{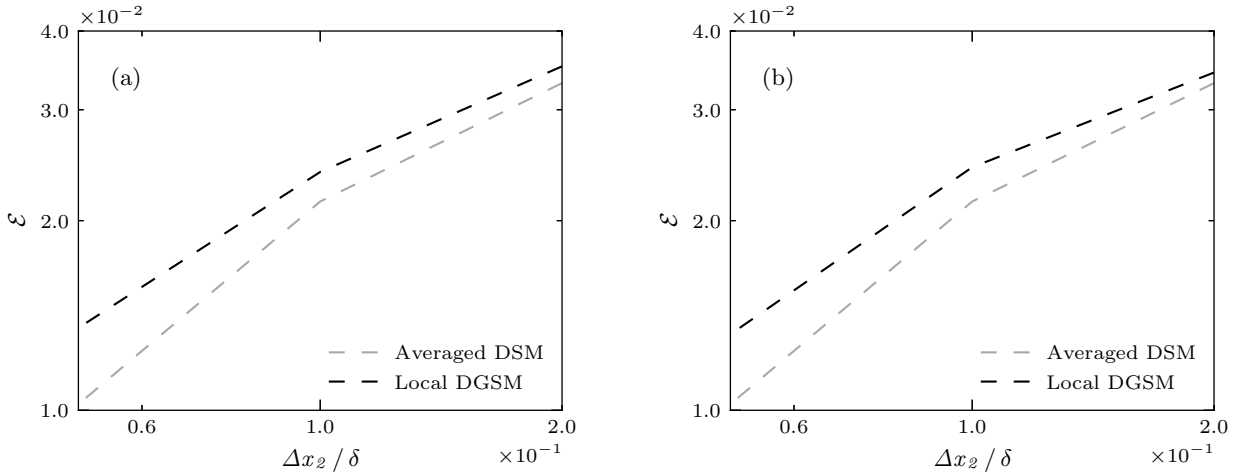}
\caption{\label{fig:wall_modeled_error} Log-log plot of the error of the mean velocity profile obtained with the averaged DSM and local DGSM as a function of grid resolution at $\textnormal{Re}_\tau \approx 4200$ (a) and $\textnormal{Re}_\tau \approx 8000$ (b).}
\end{figure}

\begin{figure*}
\includegraphics{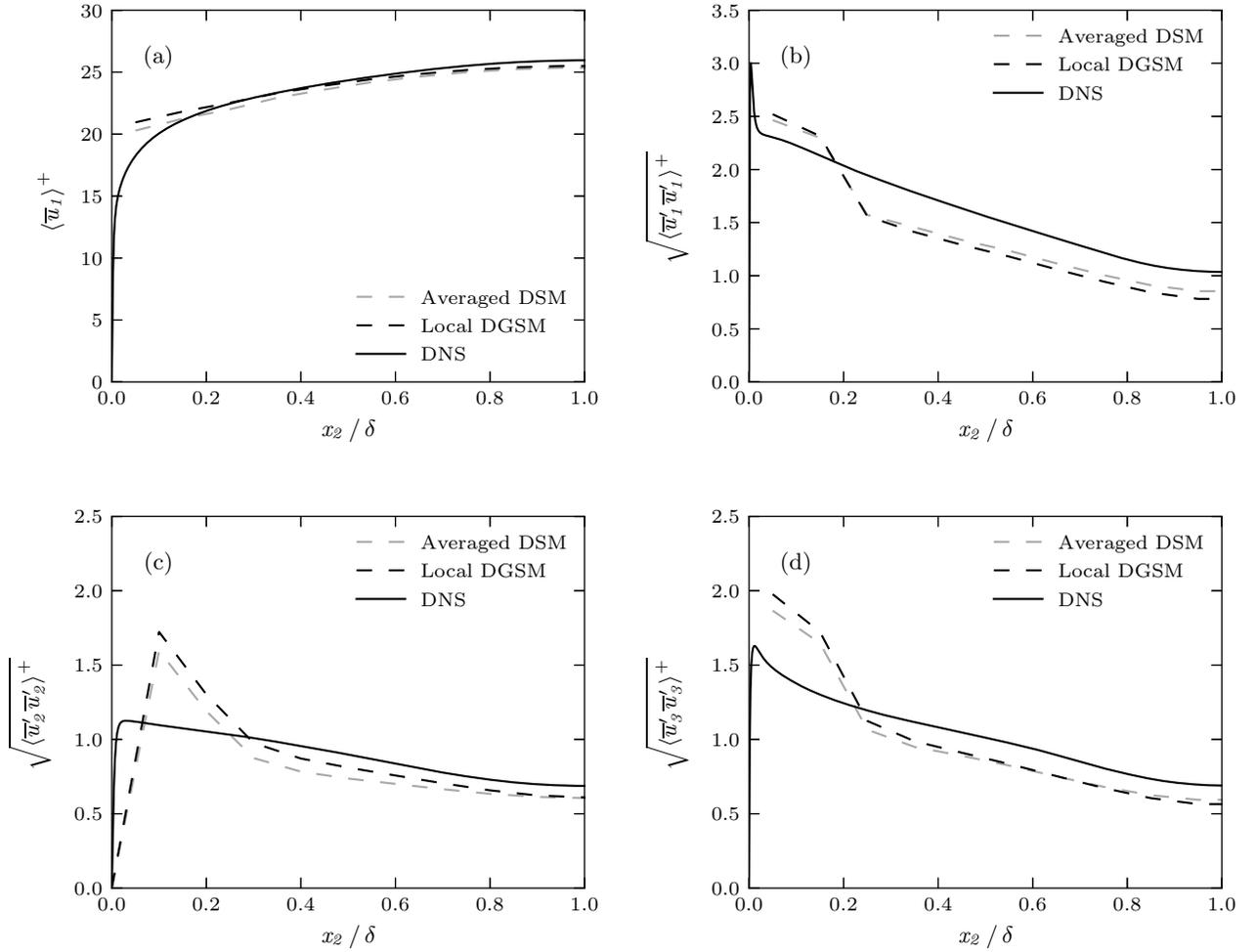}
\caption{\label{fig:chan_wall_modeled_fluctuations_grid2} The normalized mean flow velocity (a), stream-wise turbulent fluctuations (b), wall-normal turbulent fluctuations (c), and span-wise turbulent fluctuations (d) obtained with the averaged DSM and local DGSM at $\textnormal{Re}_\tau \approx 4200$ for the grid with grid spacing ${\Delta x}_2 = \delta/10$.}
\end{figure*}

The simulations with the local DGSM are stable. The error scaling of the mean flow velocity obtained with the averaged DSM and the local DGSM in the outer region for the considered grid resolutions is shown in Fig.~\ref{fig:wall_modeled_error}.
At both the considered friction Reynolds numbers, the error scales approximately linearly with the grid resolution, in agreement with the theoretical relation $\mathcal{E} \propto {\Delta x}_{2} / \delta$ for grid spacings in the inertial range \cite{lozanoduran-2019}. The errors obtained with the local DGSM are similar to errors obtained with the averaged DSM. The averaged DSM predicts the mean flow velocity profile slightly more accurately than the local DGSM. 
The mean stream-wise flow velocity and the turbulent fluctuations obtained at $\textnormal{Re}_\tau \approx 4200$ for the grid with grid spacing ${\Delta x}_2 = \delta/10$ are shown as a function of the wall-normal coordinate in Fig.~\ref{fig:chan_wall_modeled_fluctuations_grid2}. The mean flow velocity and turbulent fluctuations obtained with the averaged DSM and local DGSM closely agree. The observed underprediction of the resolved turbulent fluctuations compared to the unfiltered DNS results in the outer region is due to the missing sub-filter contributions. This is in line with expectations for wall-modeled LES and similar to results obtained in previously performed simulations \cite{bae-2018, lozanoduran-2019}.

The obtained results indicate that the accuracy of the proposed local DGSM is similar to the accuracy of the averaged DSM for wall-modeled LES of turbulent channel flow at a moderate Reynolds number.

\subsection{\label{sec:stabilitytime}Requirements for stability of simulations with local dynamic models}
The obtained LES results suggest that simulations with the local DGSM have better stability than simulations with the local DSM, and are stable without the application of averaging for time step sizes that were previously used for simulations with the AMD model. To further assess the stability of the local DGSM, the required time step size for the stability of simulations with the local DSM, local DGSM, and AMD model has been investigated for the simulations of decaying homogeneous isotropic turbulence and the temporal mixing layer. The non-dimensionalized total simulation time is set to 0.4 for decaying homogeneous isotropic turbulence and 200 for the temporal mixing layer. The time step size of simulations is decreased until the simulations are stable and do not produce unrealistic results for five consecutive smaller time step sizes. The largest of the five time step sizes for which the simulations are stable is considered to be the required time step size for stability. A larger required time step size indicates better stability of simulations with an LES model. The required time step size depends on the used numerical discretization, time-stepping method, computational grid, total simulation time, and initial condition. Therefore, the obtained required time step sizes indicate the relative stability of simulations with the different LES models but are not absolute stability requirements for general flows and simulation methods.

The AMD model locally approximates the exact sub-filter dissipation \cite{rozema-2015}. The constant of the AMD model corresponds to a spectral cutoff LES filter, whereas the local DSM and local DGSM are based on a box LES filter. To exclude the effect of differences related to the LES filter in the assessment, simulations are also performed with the AMD model with a model constant $C = 0.424$, which corresponds to a box filter \cite{rozema-2015thesis}.

The obtained required time step sizes for stability for the simulations are listed in Table~\ref{tab:stabletimestepsize}. 
\begin{table*}
\caption{\label{tab:stabletimestepsize} The required non-dimensionalized time step sizes for stability of simulations with the local DSM, the local DGSM, the AMD model, and the AMD model with the model constant corresponding to a box LES filter for the simulations of decaying homogeneous isotropic turbulence (HIT) and the temporal mixing layer on computational grids with different aspect ratios.}
\begin{ruledtabular}
\begin{tabular}{lcccc}
& Local DSM & Local DGSM & AMD model & AMD model \\
& & & & (box filter) \\
\hline
Decaying HIT & & & & \\
\hspace{3mm} Isotropic & $1.2 \times 10^{-3}$ & $2.5 \times 10^{-3}$ & $2.8 \times 10^{-3}$ & $1.9 \times 10^{-3}$  \\ %folder: /project/new_amd/cbc_timestep/. For reference: averaged DSM requires 4.2 times 10-3 (twice the AMD and DGSM tilmestep)
\hline
Temporal mixing layer & & & & \\
\hspace{3mm} Isotropic  & $8.5 \times 10^{-2}$ & $2.1 \times 10^{-1}$ & $2.1 \times 10^{-1}$ & $1.7 \times 10^{-1}$ \\ %folder: /project/new_amd/mixlayer_timestep/
\hspace{3mm} Aspect ratio 4  & $4.5 \times 10^{-2}$ & $9.5 \times 10^{-2}$ & $1.2 \times 10^{-1}$ & $8.9 \times 10^{-2}$ \\ %folder: /project/new_amd/mixlayer_v11_ar4_timestep/
\hspace{3mm} Aspect ratio 8  & $2.8 \times 10^{-2}$ & $3.4 \times 10^{-2}$ & $6.9 \times 10^{-2}$ & $4.9 \times 10^{-2}$ \\ %folder: /project/new_amd/mixlayer_v11_ar8_timestep/
\end{tabular}
\end{ruledtabular}
\end{table*}
For all the simulations the required time step size is smaller for the local DSM than for the local DGSM. The required time step size for the local DSM is on average 45.5\% smaller than the required time step size for the local DGSM. This indicates that simulations with the local DGSM have improved stability compared to simulations with the local DSM. The required time step size of the AMD model is on average 35.3\% larger than the required time step size for the local DGSM. However, the required time step size for the AMD model with the model constant corresponding to the box LES filter is on average 1.3\% smaller than the required time step size for the local DGSM. This indicates that simulations with the local DGSM have similar stability as simulations with the AMD model based on the box LES filter. Thus, the simulations with the local DGSM performed in this paper generally have similar stability compared to simulations with a static LES model which locally approximates the exact sub-filter dissipation.

\section{\label{sec:conclusion}Conclusions}
The current study has investigated the instability of simulations with the DSM without averaging the dynamic model constant and has proposed a local dynamic model for LES without averaging. It has been demonstrated that the local least-squares solution of the dynamic model constant of the DSM has a singularity. This singularity can cause exceedingly large local values of the model constant of the DSM in LES simulations, and it was hypothesized that this is one of the reasons why the DSM often requires averaging to prevent instability of simulations. The local DGSM has been proposed which removes the identified singularity by straightforwardly replacing the resolved rate-of-strain tensors in the underlying Smagorinsky model with the resolved velocity gradient tensor. These replacements were found to considerably improve the stability of the dynamic procedure, and simulations with the local DGSM are stable without averaging at time step sizes commonly used for simulations with the static AMD model. Results obtained with the DGSM for simulations of decaying homogeneous isotropic turbulence, a temporal mixing layer, and turbulent channel flow are generally as accurate as results obtained with the averaged DSM. The local DGSM can be implemented in a simulation method by making relatively simple adjustments to an existing implementation of the DSM. The good stability and accuracy, dynamic calculation of the model constant, and reduced computational complexity compared to the averaged DSM make the local DGSM a promising local dynamic model for LES.

The assessments in this study have been performed with second-order accurate simulation methods and staggered rectangular grids with limited stretching. Further assessment of the proposed local DGSM should be performed to validate its stability and accuracy for LES of complex flow, less regular computational grids, and other numerical methods. 

%\begin{acknowledgments}
% No acknowledgements to be made
%\end{acknowledgments}

%\nocite{*}

\bibliography{references}% Produces the bibliography via BibTeX.

\end{document}